\documentclass[journal]{IEEEtran}

\usepackage{url}
\usepackage{amsmath}
\usepackage[justification=centering]{caption}
\usepackage{cite}
\usepackage{pdfpages}
\usepackage{graphicx}
\usepackage{float}
\usepackage{algorithm}
\usepackage{algorithm}
\usepackage{algorithmicx}
\usepackage{algpseudocode}
\usepackage{amsmath}
\usepackage{booktabs} 
\usepackage{multirow}
\usepackage[top=2cm, bottom=2cm, left=2cm, right=2cm]{geometry}
\usepackage{setspace}
\usepackage{colortbl}
\usepackage{arydshln}
\usepackage{array}
\usepackage{subfig}
\usepackage{xcolor}
\usepackage{xspace}
\usepackage{authblk}
\usepackage{subfig}
\usepackage{bm}
\usepackage{color}
\usepackage{soul}
\usepackage{stfloats}
\usepackage{threeparttable}
\newcommand{\editorcomment}[1]{}

\renewcommand\arraystretch{1.25}
\makeatletter
\newcommand{\thickhline}{%
	\noalign {\ifnum 0=`}\fi \hrule height 1pt
	\futurelet \reserved@a \@xhline
}
\newcolumntype{"}{@{\hskip\tabcolsep\vrule width 1pt\hskip\tabcolsep}}
\makeatother

\begin{document}

\title{MCTSteg: A Monte Carlo Tree Search-based Reinforcement Learning Framework for \\Universal Non-additive Steganography}

\author{ Xianbo~Mo,
	Shunquan~Tan*,~\IEEEmembership{Senior Member,~IEEE,}
	Bin~Li,~\IEEEmembership{Senior Member,~IEEE,}
	and~Jiwu~Huang,~\IEEEmembership{Fellow,~IEEE}%
	
	\thanks{S.~Tan is with College of Computer Science and Software Engineering, Shenzhen University. X.~Mo, B.~Li, and J.~Huang are with College of Information Engineering, Shenzhen University.}%
	\thanks{All of the members are with the Guangdong Key Laboratory of Intelligent Information Processing, Shenzhen Key Laboratory of Media Security,Guangdong Laboratory of Artificial Intelligence and Digital Economy (SZ),Shenzhen Institute of Artificial Intelligence and Robotics for Society, China(email: tansq@szu.edu.cn).}%
        \thanks{*S.~Tan is the correspondence author.}%
	\thanks{This work was supported in part by the Key-Area Research and Development Program of Guangdong Province (2019B010139003), NSFC (61772349, U19B2022, 61872244), Guangdong Basic and Applied Basic Research Foundation (2019B151502001), and Shenzhen R\&D Program (JCYJ20200109105008228). This work was also supported by Alibaba Group through Alibaba Innovative Research (AIR) Program.}%
} 
\maketitle

\begin{abstract}
  Recent research has shown that non-additive image steganographic
  frameworks effectively improve security performance through
  adjusting distortion distribution. However, as far as we know, all
  of the existing non-additive proposals are based on handcrafted
  policies, and can only be applied to a specific image domain, which
  heavily prevent non-additive steganography from releasing its full
  potentiality. In this paper, we propose an automatic non-additive
  steganographic distortion learning framework called MCTSteg to
  remove the above restrictions. Guided by the reinforcement learning
  paradigm, we combine Monte Carlo Tree Search (MCTS) and
  steganalyzer-based environmental model to build MCTSteg. MCTS makes
  sequential decisions to adjust distortion distribution without human
  intervention. Our proposed environmental model is used to obtain
  feedbacks from each decision.  Due to its self-learning
  characteristic and domain-independent reward function, MCTSteg has
  become the first reported universal non-additive steganographic
  framework which can work in both spatial and JPEG domains. Extensive experimental results show that MCTSteg can effectively
  withstand the detection of both hand-crafted feature-based and
  deep-learning-based steganalyzers. In both spatial and JPEG domains,
  the security performance of MCTSteg steadily outperforms the state
  of the art by a clear margin under different scenarios.
\end{abstract}


\bstctlcite{psm_bstctl}

\IEEEpeerreviewmaketitle

\section{Introduction}

\IEEEPARstart{S}{teganography} hides secret information into stego media and try to evade detection from steganalysis, where spatial and JPEG domain images are the most common cover media~\cite{li2011survey}. Almost all modern steganographic algorithms are based on a distortion minimization framework~\cite{fridrich2007practical}, decomposing a steganographic algorithm into the design of cost function and coding scheme. When we study the cost function, the optimal embedding simulator can be used to simulate real embedding impact. Started by Pevn\'y and Filler \cite{pevny2010using}, several heuristic cost functions such as \cite{Vojt2012Designing, Vojt2014Universal, 2015A, 2017A, 6776485, 6607427}have been proposed.

These algorithms attain global distortion by summing up the local costs of individual pixels and thus were called \emph{additive steganography}. However, the application of content-adaptive cost function causes embedding modifications clustering in textured areas, so inter-pixel correlations and interactions among these modifications destroy the prerequisite for additive distortion. Filler and Fridrich~\cite{2010Gibbs} introduced \textit{non-additive steganography} in the spatial domain. Later, distortion adjustment strategies were proposed to guide the direction of embedding modifications. In CMD(Clustering Modification Directions)~\cite{2015CMD} and Synch~\cite{Denemark2015Improving}, the authors pointed out that clustering modification directions are helpful to improve security performance. In the JPEG domain, Li et al.~\cite{li2018defining} proposed a strategy called BBC(Block Boundary Continuity) to maintain the block boundary continuity. Wang et al.~\cite{9144531} first developed a non-additive framework called BBC++ based on BBC and then proposed BBM(Block Boundary Maintenance)~\cite{wang2020non} to minimize the modifications on the spatial block boundaries, which can be combined with BBC. As for coding schemes~\cite{2007Improving, 2010Near, 2011Minimizing}, STC(Syndrome–Trellis Codes) is the most commonly used one for practical application. Recently, based on polar codes, Zhang et al.~\cite{9044194} proposed a better near-optimal steganographic coding method called SPC(Steganographic Polar Codes).

Steganalysis detects secret bits hidden in cover media. The most well-known traditional steganalyzers are the ``rich models''~\cite{Fridrich2012Rich,2014Selection,2016Adaptive,2017Pixel,2017New}, a hand-crafted feature family, equipped with ensemble classifier~\cite{2012Ensemble}. Recently, this battleground has been dominated by deep learning frameworks. Based on auto-encoders, Tan and Li~\cite{2014Stacked} made the first attempt, and then Xu et al.~\cite{2016Structural} proposed a deep learning-based network with convolutional and linear layers, which is a milestone in steganalysis networks. Later, deeper and more complex structures\cite{qian2015deep,yedroudj2018yedroudj,2017Deep,10.1145/3082031.3083248,10.1145/3082031.3083236,8125774} were proposed for either spatial- or JPEG-domain by exploiting background knowledge of hand-crafted feature based steganalysis. Boroumand et al.~\cite{2018Deep} proposed SRNet, a domain-independent deep residual network with superior detection performance, whose objective is minimizing the use of heuristic domain knowledge. Tan et al.~\cite{9126847} proposed CALPA-NET, a channel pruning-assisted deep residual network architecture to shrink the size of existing vast and over-parameterized deep learning-based steganalyzers. Yousfi et al.\cite{yousfi2020imagenet} and Butora et al.\cite{butora2021pretrain} successfully refined the ImageNet\cite{deng2009imagenet} pretrained computer vision models, e.g. EfficientNet\cite{tan2019efficientnet}, for steganalysis tasks with similar or better performance compared with SRNet.

Meanwhile, deep learning structure has also been applied in steganographic methods.  ASDL-GAN~\cite{2017Automatic} and UT-GAN~\cite{yang2019embedding} are the most representative deep learning-based steganographic frameworks. Both ASDL-GAN and UT-GAN are composed of two subnetworks, the first one is steganographic generative subnetwork, which aims at learning embedding probabilities. The second one is steganalysis discriminative subnetwork, which tries to distinguish whether the input is cover or stego. Moreover, based on the reinforcement learning paradigm, Tang et al.~\cite{tang2020automatic} proposed Steganographic Pixel-wise Actions and Rewards with Reinforcement Learning (SPAR-RL), which aims at maximizing the rewards evaluated by the steganalysis environment. Please note that, all these three mentioned frameworks are additive steganographic framework, thus their security performance is far behind non-additive steganographic frameworks. But, on the other hand, there are defects in current non-additive steganography\cite{2015CMD, Denemark2015Improving, li2018defining, 9144531, wang2020non}. For example, all of them are based on handcrafted distortion adjustment strategies. Furthermore, these strategies are designed for specific domain, implying that they can only work in either spatial or JPEG domain. \par
In this paper, we develop a novel universal non-additive steganographic framework called \emph{MCTSteg}, aiming at automatically adjusting distortion without human intervention. This framework is inspired by the huge success of the AlphaGo family~\cite{2016Mastering, 2017Mastering}, the first computer program to defeat a professional human Go player, because we think there are similarities between Go and non-additive steganography. For steganographers, they modify cover media by $\pm 1$ to transmit secret message. Specially, in non-additive steganography, steganographers control the location distribution of $\pm 1$ modifications by adjusting embedding distortion, which is similar to put down black and white pieces on a Go board. As the core component of AlphaGo infrastructure, the Monte Carlo Tree Search (MCTS), a robust machine learning method, is utilized to find optimal solutions in given space by building a search tree. Therefore, we adopt MCTS tree into our framework. From the respect of game theory, MCTSteg is composed of two modules, where the deep learning-based steganalyzer acts as environmental model and MCTS tree acts as non-additive steganographer. Guided by background knowledge, we define the reinforcement learning elements to connect MCTS and steganography. To win the battle between steganographer and steganalyzer, MCTSteg first divides the cover image into several sublattices. For each sublattice, its distortion distribution of different modification polarities is adjusted according to the search result of MCTS without any non-additive steganographic rules. For the adjustment order, we design a new distortion-adaptive strategy called \emph{distortion descending order (DDO)}. Next, we design a reward function to calculate feedbacks of each decision made by MCTSteg. Because the reward function can work in both spatial and JPEG domain, MCTSteg is a native universal framework. After executing enough searches, MCTSteg can learn to generate more secure embedding distortion. Extensive experiments conducted on three datasets with the optimal embedding simulator show that under the detection of both hand-crafted feature-based and deep learning-based steganalyzers, stego images generated by MCTSteg achieve the best statistical score and security performance compared with state-of-the-art non-additive steganographic frameworks and  machine learning-based steganography.

The rest of this paper is organized as follows. In Sect.~\ref{SecPre}, we make a brief overview of the preliminaries in our research. Then we describe the technical roadmap and challenges of MCTSteg in Sect.~\ref{SecMCTSteg}. To demonstrate the effectiveness of our framework, Sect.~\ref{SecExp} first presents experimental setup and statistical analysis. Next, it compares security performance against various steganalyzers with state-of-the-art steganographic methods. Finally, we conclude and list our future work in Sect.~\ref{SecConclu}.

\section{Preliminaries}\label{SecPre}

\subsection{Non-additive Steganographic Framework}\label{SecNSF}

Started by Filler and Fridrich, they proposed the Gibbs construction framework~\cite{2010Gibbs} which consists of four steps. Firstly, local potentials are calculated by the cost function so that the distortion minimization framework can be employed. The distortion~$\boldsymbol{D}$ that cover$~\boldsymbol{C}$ changes into stego$~\boldsymbol{S}$ is expressed as:
\begin{equation}
	D(C,S) = \sum_{i=1}^{r}\sum_{j=1}^{l}\left[{\rho_{i,j}^+}\delta(d_{i,j}-1)+{\rho_{i,j}^-}\delta(d_{i,j}+1)\right],
\end{equation}
where~${{\rho^{+}_{i,j}}}$ and~${{\rho^{-}_{i,j}}}$ are the distortions that~$c_{i,j}$ modifies by +1 and -1 respectively; $d_{i,j} \in \boldsymbol{D}$ represents the difference between the stego and the cover; the dimension of $\boldsymbol{C,S,D,\rho^+,\rho^-}$ is all$~r\times~l$, which is decided by the dimension of $\bm{C}$. $\delta(\cdot)$ is an indicator function:
\begin{equation}
	\delta(x)=
	\begin{cases}
		1 & x=0      \\
		0 & x\neq 0.
	\end{cases}
\end{equation}
Secondly, the cover image is decomposed into disjoint sublattices with a distance larger than the support width of local potential. Specially, in JPEG-domain, sublattices are divided based on 8$\times$8 DCT block. Thirdly, in each iteration, one sublattice is embedded while the distortion of others are updated. Finally, the iterations are repeated until convergence, and the introduced embedding pattern is expected to be a sample from optimal embedding.

Prior research, including CMD~\cite{2015CMD} and Synch~\cite{Denemark2015Improving}, has demonstrated that based on the additive cost function, synchronizing the direction of embedding modification improves security performance with a small entropy of stego noise. In CMD, modification synchronization is achieved by adjusting the distortion contribution as:

\begin{IEEEeqnarray}{rCl}
\rho_{i,j}^+&=&
\begin{cases}
	\rho_{i,j}^+/\alpha, & \text{If nearby modification is more than +1} \\
	\\
	\rho_{i,j}^+,        & \text{otherwise}
\end{cases}\label{rho_plus}\\
\rho_{i,j}^-&=&
\begin{cases}
	\rho_{i,j}^-/\alpha, & \text{If nearby modification is less than -1} \\
	\\
	\rho_{i,j}^-,        & \text{otherwise},
\end{cases}\label{rho_minus}
\end{IEEEeqnarray}

where~$\alpha$ is a scaling factor. A larger~$\alpha$ leads to a more concentrated embedding direction.

\subsection{Monte Carlo Tree Search}\label{SecMCTS}

When we face the challenge of tremendous search space dimensionality such as Go, MCTS is always robust to make the optimal decisions by building an MCTS tree with its search results. The whole procedure is composed of four stages:

\begin{enumerate}
	\item \emph{Selection}:
	      Starting from the root node, the selection direction is guided by the tree policy. Each time only one node is chosen from the available child node set. During the whole search process, the visiting counts of all selected nodes are increased by one. Finally, if a fully expanded node is not reached, the search procedure goes to \textit{expansion}. If a leaf node is reached, the search procedure goes to \textit{backpropagation}.
	\item \emph{Expansion}:
	      An unvisited child node is randomly selected and added to the MCTS tree. Then it goes to \textit{simulation}.
	\item \emph{Simulation}:
	      A default policy such as random search guides the child node selection and updates parameters until a leaf node is reached. Thus the search result is obtained, and the search procedure goes to \textit{backpropagation}.
	\item \emph{Backpropagation}:
	      The reward or penalty feedback of the search result is calculated and added to the cumulative reward of each selected node in the search path.
\end{enumerate}
The most popular \textit{tree policy} is \emph{Upper Confidence Bound for Trees}~\cite[UCT]{auer2002finite}. Let~$v$ denote a child node and~$v'$ be its parent. The UCT score of~$v$ is calculated by:
\begin{equation}
	\text{UCTScore}(v) = \frac{R(v)}{N(v)} + C\sqrt{\frac{\ln N(v)}{N(v')}},
\end{equation}
where~$R(\cdot)$ represents the cumulative reward, $N(\cdot)$ is the visiting count. The two terms in this equation make a balance between exploration and experience where~$C$ is a weighting factor. As a strategy, UCT selects the node with the maximum UCT score. However, it relies on the statistics of visited nodes, which means that it cannot be applied to unvisited nodes. Therefore, in \textit{expansion} and \textit{simulation}, random selection is adopted as the default policy. With a lower complexity, the search procedure can be executed faster so that more results can be generated in a limited time.

\begin{figure*}[!h]
	\centering
	\includegraphics[scale=0.77]{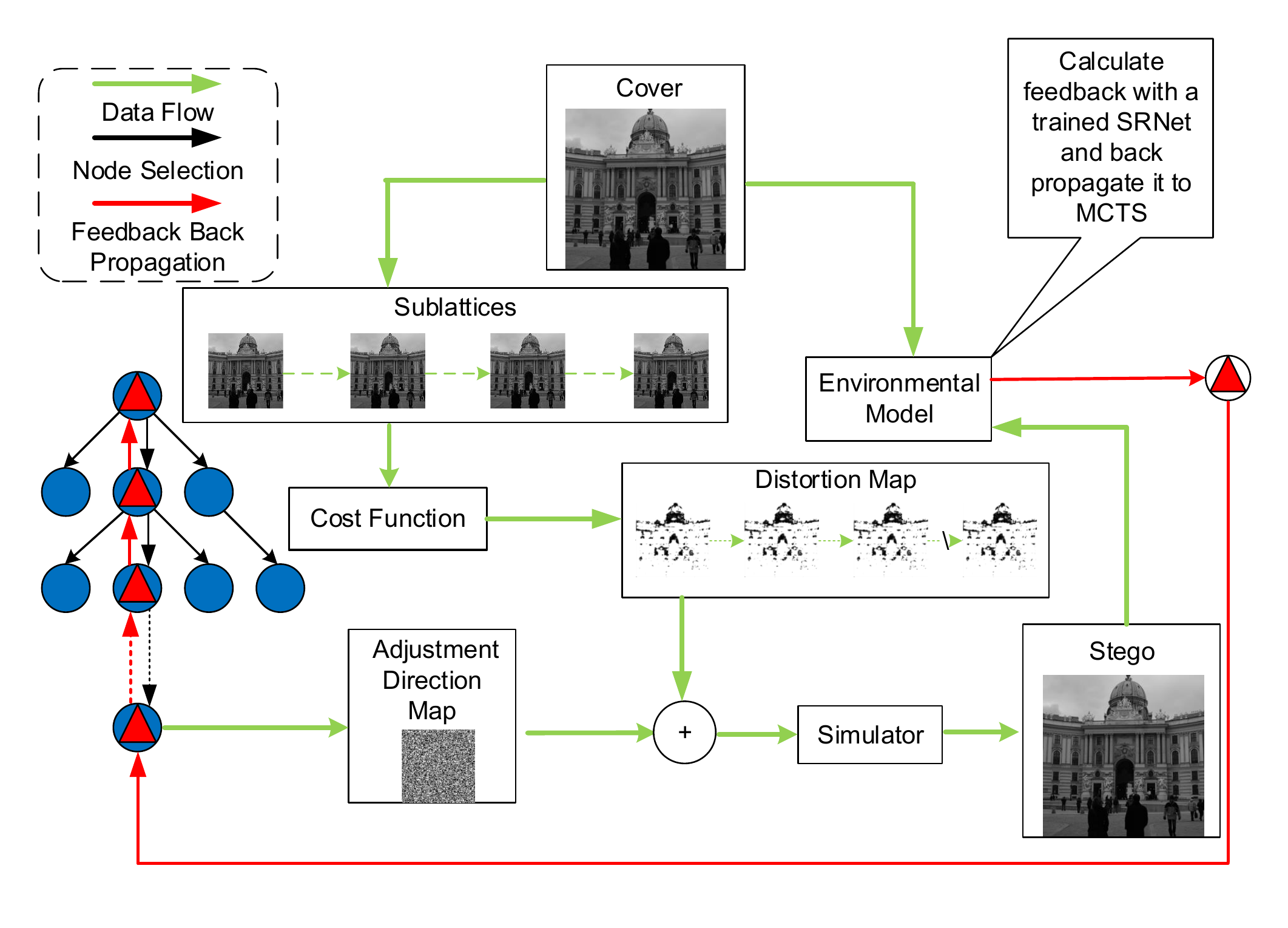}
	\caption{The overall structure of MCTSteg. It first divides cover image into four sublattice, then embeds and updates them in turns according to distortion which has been adjusted by the search result of MCTS tree. Later MCTSteg will calculate feedback with the help of environmental model and back propagate it to nodes in the search path. After all sublattices have been embedded, the stego image generated by MCTSteg can be obtained.}
	\label{fig2}
	
\end{figure*}

\subsection{Reinforcement Learning}\label{Secreinforcement learning}
Reinforcement learning is a self-learning algorithm and can be used to develop game strategies through continuous interaction between players and environments. To model the real scenario, reinforcement learning contains the following basic elements:
\begin{itemize}
	\item $S$: A set of states, where~$s^{0}$ denotes the initial state and~$s^{T}$ denotes the terminal state.
	\item $A$: A set of actions. For each~$s^{i} \in S$, an action~$a$ is sampled from~$A$ and transfer~$s^{i}$ to~$s^{i+1}$.
	\item $f(s,a,s')$: A function that denotes the transfer from~$s$ to~$s'$ with action~$a$.
	\item $\pi$: A policy~$\pi$ is a probability distribution that guides the action selection.
	\item $R(s^{i}, a, s^{i+1})$: The reward value of selecting action~$a$ at state~$s_{i}$.
	\item $E$: An environmental model for evaluating~$s^T$.
\end{itemize}

As a game-like situation, players have to make a series of decisions from~$s^0$ to~$s^T$ based on~$\pi$. Then~$R$ is calculated by~$E$ and assigned to all related actions. The goal of reinforcement learning is to achieve the maximum total reward. In Sect.~\ref{SecMCTSteg}, we present more details of reinforcement learning elements used in our proposed framework.

\section{Proposed Framework}\label{SecMCTSteg}

\subsection{Technical Roadmap and Challenges}\label{SecTech}
In game theory, most games can be modeled by basic elements including players, environmental models, states, actions, and rewards. Taking Go as an example, starting from the initial state of a blank Go board, players elaborately take actions until the result is settled down according to Go rules which can be seen as an environmental model.

For the game between non-additive steganographer and steganalyzer, the embedding distortion calculated by the additive cost function is our initial state. To win this game, the non-additive steganographer takes a series of actions to adjust the distortion distribution. Then stego images are generated according to the adjusted distortion, and the steganalyzer classifies these images. If the steganalyzer gives a wrong prediction, the steganographer wins this game. Otherwise, the steganographer loses. Therefore, this well-trained steganalyzer acts as an environmental model. As we all know, MCTS is a wise decision-maker~\cite{browne2012survey}, but it is doubtful whether MCTS can be a good steganographer without human guidance. To address this issue, we build an MCTS-based non-additive steganographic framework. There are three major challenges:
\begin{enumerate}
	\item \emph{Relevance}:
	      MCTS is a tool for making decisions in a search space, while steganography is an information hiding technique. It is difficult to combine the search space of MCTS with the distortion metric of the distortion minimization framework.
	\item \emph{Game Rules}:
	      In most games such as Go and Chess, the game rules are clearly defined, so the game results can be easily obtained. In steganography, there are no such rules available.
	\item \emph{Complexity}:
	      The initial state in non-additive steganography is different among cover images because their distortion distributions vary, which means that we have to discuss individual distortion adjustment for each of them.
\end{enumerate}

As a consequence of these difficulties, the goal of our proposed framework MCTSteg is not discovering the optimal distortion adjustment but defeating the well-trained environmental model, which is similar to adversarial examples~\cite{nguyen2015deep}. Moreover, the game results are decided by the classification of environmental model, which solves the problem of game rules. However, considering the computational complexity, we do not use gradient to attack the environmental model but adapt the reinforcement learning theory and design a low-complexity reward function. The details of our framework will be discussed in  the rest of Sect. \ref{SecMCTSteg}.
\begin{figure*}[!htp]
	
	\centering
	\includegraphics[scale=0.421]{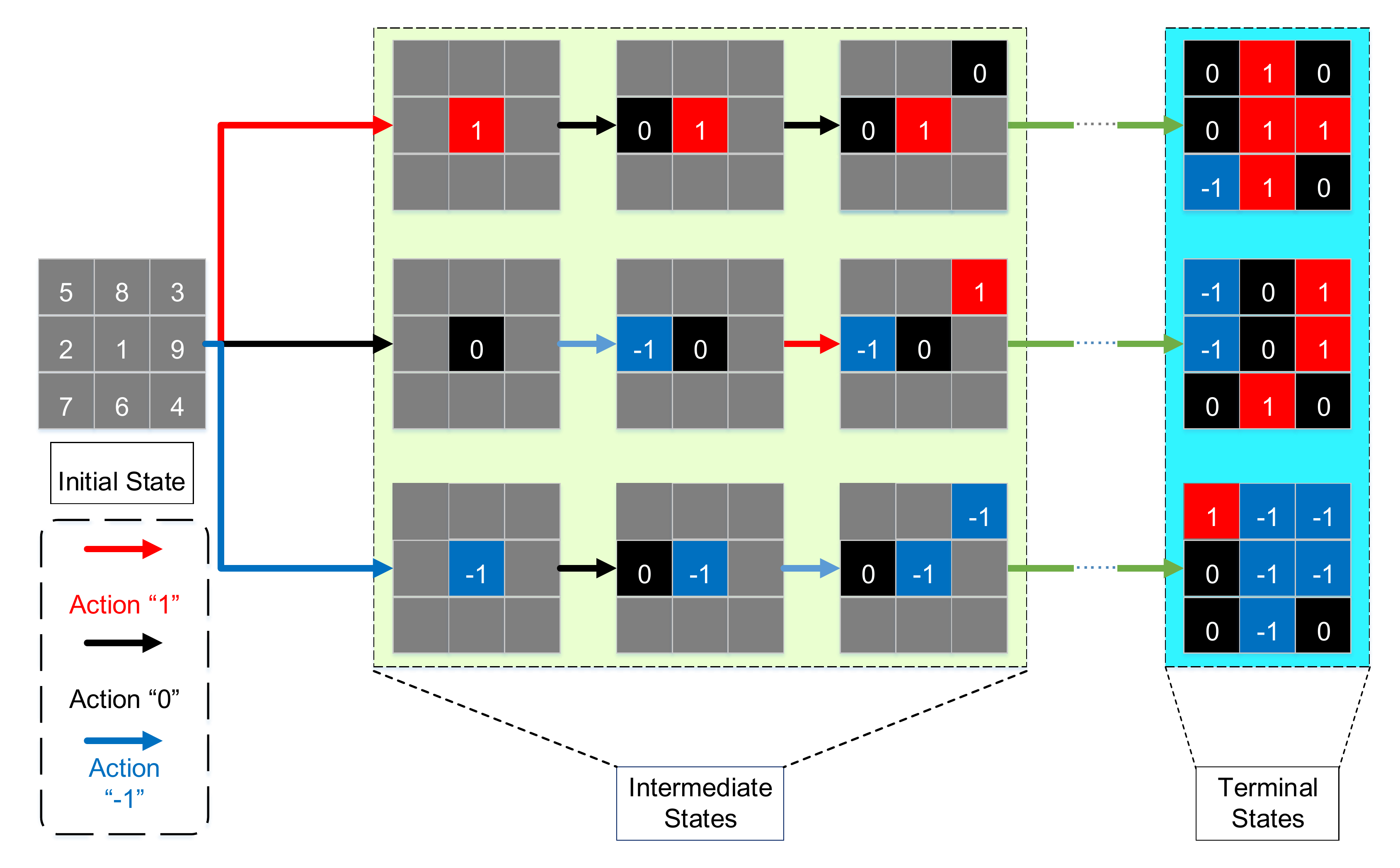}
	\caption{An example of state transfer. The numbers in the initial state are distortion adjustment order. The red arrow represents for action ``1'' while black arrow is for ``0'' and blue arrow is for ``-1''. An action is taken means that the value of corresponding elements of $\Gamma$ will be changed.}
	\label{fig3}
	
\end{figure*}
\subsection{Element Definitions}\label{SecDef}
Let~$\bm{\rho^{-}}$ denote the embedding distortion matrix of~$-1$ modifications and~$\bm{\rho^{+}}$ for~$+1$ modifications. Equations~\ref{rho_plus} and~\ref{rho_minus} can be expressed in the matrix form:
\begin{eqnarray}
	\bm{\rho^{+}}\prime &= \bm{\rho^{+}}\bm{\omega^{+}} \\
	\bm{\rho^{-}}\prime &= \bm{\rho^{-}}\bm{\omega^\bm{-}},
\end{eqnarray}
where~$\bm{\omega^{+}}$ and~$\bm{\omega^{-}}$ are the distortion adjustment coefficient matrices for~$\bm{\rho^{+}}$ and~$\bm{\rho^{-}}$ respectively, $\bm{\rho^{+}}\prime~and~\bm{\rho^{-}}\prime~$ are the adjusted distortion matrices, the mathematical operation is dot multiplication. To better control the values in~$\bm{\omega^{+}}$ and~$\bm{\omega^{-}}$, we define a distortion adjustment polarity matrix~$\bm{\Gamma}=(\gamma_{i,j})^{r\times l} $, where~$r$ and~$l$ are the dimensions and~$\gamma_{i,j} \in \{0,+1,-1\}$ in ternary embedding. Therefore,~$\bm{\omega^{+}}$ and~$\bm{\omega^{-}}$ are expressed as:

\begin{eqnarray}
	\bm{\omega^{+}}_{i,j} &=&
	\begin{cases}
		1/\alpha,       &~\text{If}~ \gamma_{i,j} = +1 \\
		1, & \gamma_{i,j} = 0
	\end{cases}\label{equ:omega_plus}\\
	\bm{\omega^{-}}_{i,j} &=&
	\begin{cases}
		1/\alpha,       &~\text{If}~ \gamma_{i,j} = -1 \\
		1, & \gamma_{i,j} = 0.
	\end{cases}\label{equ:omega_minus}
\end{eqnarray}

As~$\bm{\Gamma}$ decides distortion adjustment, we take MCTS into consideration because it can search the optimal sample in the distribution space of ~$\bm{\Gamma}$. Before presenting the search process, we first define the following reinforcement learning elements to build the theoretical basis of MCTS tree:
\begin{itemize}
	\item \emph{States}:
	      The set of states is denoted by~$\bm{\Gamma}$, which acts as parameter of tree nodes in MCTS. Specifically, we define~$\bm{\Gamma}^{0}$ for the initial state as:
	      \begin{equation}
		      \bm{\Gamma}^{0} = \left\{ \gamma_{i,j}=0 \,|\, \forall (i,j) \in \{1,\ldots,r\} \times \{1, \ldots, l\} \right\},
	      \end{equation}
	      where~$\bm{\rho^{+}}$ equals~$\bm{\rho^{-}}$ at the initial state. We also define~$\bm{\Gamma}^{T}$ for the terminal state, where $\{ \gamma_{i,j}\,|\, \forall (i,j) \in \{1,\ldots,r\} \times \{1, \ldots, l\} \}$ have been assigned according to the search result of MCTS. An example procedure of transferring from the initial state to the terminal state is illustrated in Fig.~\ref{fig3}.

	\item \emph{Actions}:
	      An action~$a$ takes one value in~$\{-1,0,+1\}$. Taking an action~$a$ means that the value of~$\gamma_{i,j}$ will be changed to~$a$ based on selection policies, corresponding to node selection in MCTS. The illustration of action is also shown in Fig.~\ref{fig3}.

	\item \emph{Feedbacks}:
	      As a zero-sum game, the feedback in non-additive steganography can only be determined until we attain~$\bm{\Gamma}^{T}$. The most common way to obtain feedbacks of intermediate states is using the same feedback of~$\bm{\Gamma}^{T}$. To evaluate the security performance of samples generated by MCTSteg, we design a reward function using the classification confidence. Let~$f(\cdot)$ denote the well-trained environmental model and~$f_c(\cdot)$ represent the confidence of~$f(\cdot)$ to classify the input as cover. The reward feedback~$R$ is calculated by:
	      \begin{equation}\label{equ:reward}
		      R = f_c(M) - f_c(Y),
	      \end{equation}
	      where~$M$ is the sample generated by MCTSteg,~$Y$ is the stego image used to attain security performance baseline. $f(\cdot)$ can roughly reflect the security performance of its inputs. Therefore if~$M$ can more effectively resist against the environmental model (i.e.,~$R$ is positive),~$M$ will achieve better security performance compared with~$Y$. If not (i.e.,~$R$ is negative),~$M$ is worse. Theoretically, the better the environmental model we adopt, the values of~$R$ is closer to its actual distribution, which can help MCTSteg achieve better security performance. As mentioned in \cite{Kartal_aaai_2019}, the challenge of sample inefficiency is widely existed in the MCTS application. In our paper, we have also found that for the majority of samples generated by MCTSteg, their rewards are negative while the goal of MCTSteg is to achieve maximal total reward. To improve sample efficiency, researchers have proposed various methods in the past. In our proposed MCTSteg, we adopt a simple solution, that is, we assign the samples with positive $R$ a higher reward to encourage MCTSteg to generate more positive samples:
	      \begin{equation}\label{equ:scalefactor}
	      	R' = 
	      	\begin{cases}
	      		R*10,       &~\text{If}~ R>=0 \\
	      		R*1, &~\text{otherwise},
	      	\end{cases}
	      \end{equation}
      	where the $R'$ is the scaled feedback, $10$ and $1$ are the scaling factors corresponding to positive and negative samples respectively.
\end{itemize}

In Sect.~\ref{SecOverall}, we will present how to utilize these elements defined above to build MCTSteg.

\subsection{Overall Framework}\label{SecOverall}
Our proposed MCTSteg is composed of MCTS and a steganalyzer-based environmental model. The MCTS adjusts embedding distortion while the environmental model outputs the corresponding feedback introduced in Sect.~\ref{SecDef}. The training of MCTSteg is similar to learn to defeat the environmental model. To do this, MCTS has to develop its strategies based on the corresponding environmental model's feedback and try to attain global optimal search result.

MCTS has a trigeminal-tree structure, which is decided by the number of action types. The tree node has seven parameters:
\begin{itemize}
	\item \emph{n}: Visiting counts
	\item \emph{r}: Cumulative reward
	\item \emph{p}: Parent node
	\item \emph{lc}: Left child
	\item \emph{mc}: Middle child
	\item \emph{rc}: Right child
	\item \emph{d}: Adjustment order
	\item \emph{$\bm{\Gamma}$}: Distortion adjustment polarity matrix
\end{itemize}
\begin{figure}[!htp]
	
	\centering
	\hspace{-1cm}
	\includegraphics[scale=0.40]{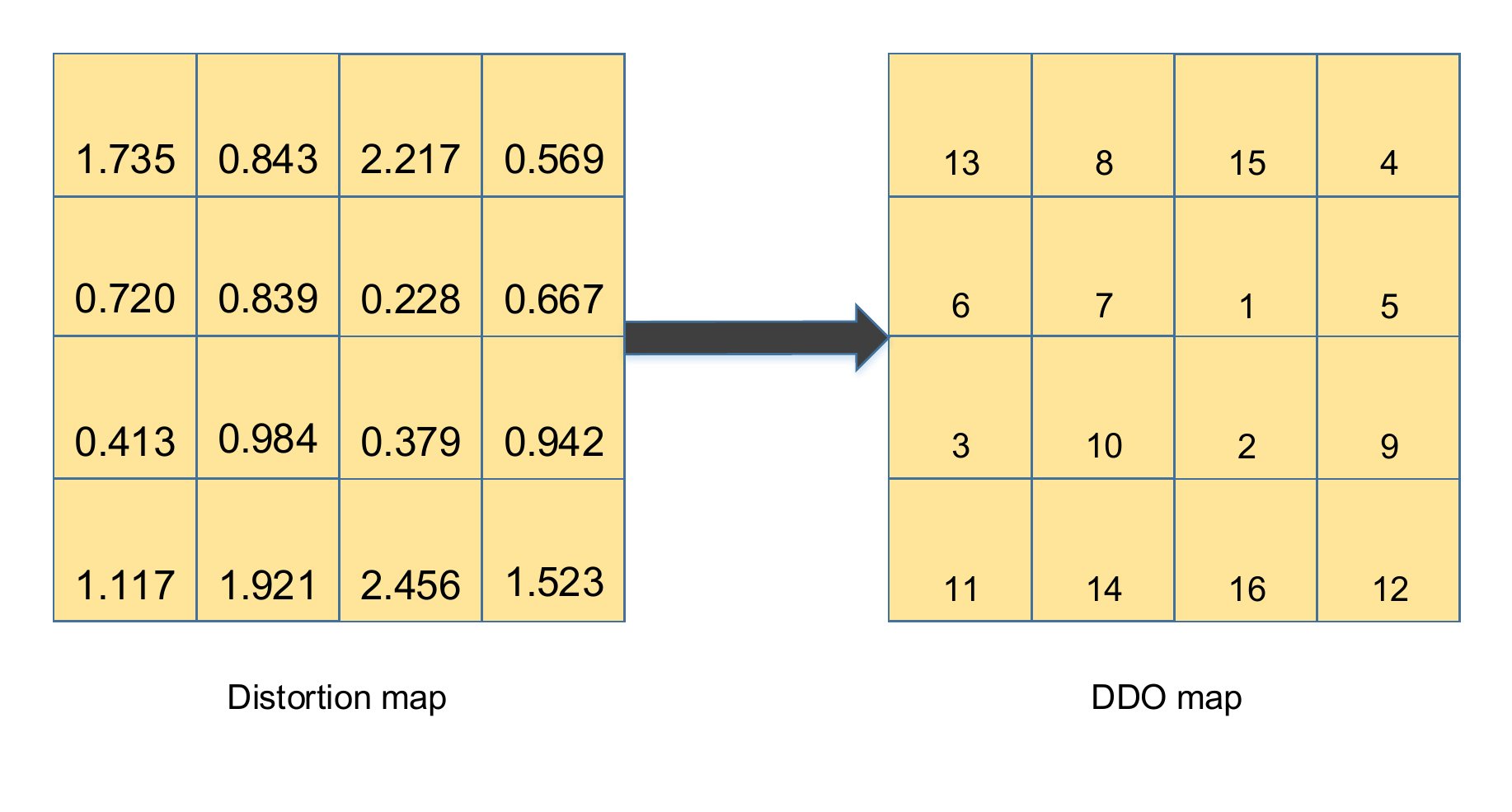}
	\caption{The left matrix represents for a distortion map and the elements of DDO map is the descending order of distortion map. }
	\label{fig6}
	
\end{figure}
\par
During the search process, MCTSteg will sequentially make decisions until attaining terminal state. Each decision decides the distortion adjustment polarity of one specific pixel. Meanwhile, the nodes of MCTS in upper layers are visited more frequently, so their statistics are closer to actual distributions. Therefore, we design a new adjustment strategy called \textbf{distortion descending order (DDO)}. Firstly the pixels are sorted by their cost in descending order. Then we adjust distortion in that order. For example, in MCTSteg, pixels with lower cost are processed in upper layers while pixels with higher cost are processed in lower layers. Thus MCTSteg is more likely to process the pixels with lower cost, which effectively improves security performance.

Based on non-additive steganography, we design Algorithm~\ref{Algo1} and Algorithm~\ref{Algo2}, which can be summarized as following workflow:

\begin{algorithm}
	\caption{Modules for MCTSteg}
	\label{Algo1}
	\begin{algorithmic}[1]
		
		\Function{GetCoordinate}{$d$}
		\State $x \gets d~/~l,~y \gets d \bmod l$
		\State return $(x,y)$
		\EndFunction
		
		\Function{BestChild}{$v$}
		\State $C \gets \text{child set of}~v$
		\State $v' \gets \mathop{\arg\max}_{c \in C} \text{UCTScore}(c)$
		\State $v'.n \gets v'.n+1$
		\State return $v'$
		\EndFunction
		
		\Function{Backpropagate}{$v$}
		\State Calculate~$R$ using Equation~\ref{equ:reward}
		\While{$v$ is not root node}
		\State $v.r \gets v.r + R,~v \gets v.p$
		\EndWhile
		\EndFunction
		
		\Function{RandomSearch}{$v$}
		\While{$v$ is not a leaf node}
		\State Randomly select an untried action~$a \in A(v)$
		\State Create a new node~$v'$
		\State $(x,y)\gets $\Call{GetCoordinate}{$v'.d$}
		\State $v'.p\gets v,~v'.d\gets v.d+1,~v'.\bm{\Gamma}\gets v.\bm{\Gamma}$
		\State $v'.\gamma_{x,y}\gets a$
		\If{$a=-1$}
		\State $v.lc \gets v'$
		\ElsIf{$a=0$}
		\State $v.mc \gets v'$
		\Else
		\State $v.rc \gets v'$
		\EndIf
		\State $v \gets v'$
		\EndWhile
		\State return $v$
		\EndFunction
		
		\Function {Search} {$v$}
		\While{$v$ is fully expanded}
		\State $v \gets $ \Call{BestChild}{$v$}
		\EndWhile
		\If{$v$ is not a leaf node}
		\State $v \gets $ \Call{RandomSearch}{$v$}
		\EndIf
		
		\State return $v$
		\EndFunction
	\end{algorithmic}
\end{algorithm}

\begin{enumerate}
	\item \emph{Setup}: Let~$X$ denote the cover image. MCTSteg initializes the stego image~$Y = X$, decomposes~$Y$ into~$N$ sublattices, and divides the embedding message into~$N$ segments. Let~$S_t$ denote the~$t$-th sublattice,~$M_t$ denote the~$t$-th segment, and~$Y_t$ denote partially embedded~$Y$ where $S_1,\ldots,S_{t-1}$ have been embedded. MCTSteg embeds~$S_1$ with~$M_1$ and goes to \emph{Distortion Update} with~$t=1$.

	\item \emph{Distortion Update}: MCTSteg increases~$t$ by one. If~$t>N$, it outputs~$Y_t$ as final stego images. Otherwise, it computes distortion based on~$Y_{t-1}$ with a cost function such as HILL and S(J)-UNIWARD and goes to \emph{Sample Selection}.

	\item \emph{Sample Selection}: MCTSteg runs \emph{Distortion Adjustment} within the computational budget and selects the~$Y_{t}^{\text{tmp}}$ with the highest~$R$ as~$Y_t$.

	\item \emph{Distortion Adjustment}: MCTSteg executes the \textit{search function} in Algorithm~\ref{Algo1} on~$S_t$ and obtains the distortion adjustment polarity matrix~$\bm{\Gamma}$. It then embeds~$M_t$ into~$S_t$ according to the distortion adjusted by~$\bm{\Gamma}$, updates temporary stego~$Y_{t}^{\text{tmp}}$ with~$S_t$, and executes the \textit{backpropagate function} in Algorithm~\ref{Algo1}.

\end{enumerate}

Inside the \textit{search function}, we adopts UCT strategy for nodes that have been fully expanded (i.e., all of three child nodes including left child, middle child, and right child have been visited) and random search for nodes that have not been fully expanded. For \textit{backpropagate function}, MCTSteg will recursively accumulate the feedback to each node in the search path.

In Algorithm~\ref{Algo2}, the computational budget is designed based on the goal of the reinforcement learning paradigm, which is to achieve the highest total reward. The reward value is decided by~$R$ of every search result and the total search times. Therefore, we set up two computational budgets. The first computational budget considers the time cost. We set a maximum search count for each sublattice. If the number of \textit{Distortion Adjustment} executions reaches the maximum search count, we choose the~$Y_{t}^{\text{tmp}}$ with the highest~$R$ as the globally optimal result for~$S_t$. The second one corresponds to~$R$. If~$Y_{t}^{\text{tmp}}$ is classified as cover by the environmental model with confidence over the confidence threshold, then MCTS is considered to have found the highest~$R$ corresponding to the globally optimal result for~$S_t$. It is worthless to continue search so we can stop \textit{Distortion Adjustment} in advance. We think that the proper confidence threshold can both ensure the sufficient security performance of final search result and avoid wasting superabundant time cost in search procedure. In Sect. \ref{SecHyper}, we will conduct experiments to discover the proper confidence threshold.

\begin{algorithm}
	\caption{MCTSteg Algorithm}
	\label{Algo2}
	\begin{algorithmic}[1]
		\Require Cover~$X$, Message~$D$
		\Require Environmental Model~$R$
		
		\Function{Main}{$X$, $M$, $R$}
		\State $Y \gets X,~ t \gets 1,Y_t \gets Y$
		\State Compute~$\bm{\rho^{+}}$ and~$\bm{\rho^{-}}$ with the cost function
		\State $Z \gets\Call{Simulator} {X, \bm{\rho^{+}}, \bm{\rho^{-}}, M}$
		
		\State $S_1\gets\Call{Simulator} {S_1, \bm{\rho^{+}}, \bm{\rho^{-}}, M_1}$
		\State Update~$Y_2$ with~$S_1$
		\While{$t \leq N$}
		\State $t \gets t+1$
		\State Create root node~$V^0$ with state~$\bm{\Gamma^{0}}$
		\State Get sublattice~$S_t$ from~$Y$ and~$M_t$ from~$M$
		\State $R_{\text{top}} \gets 0$, $Y^{\text{tmp}}_t \gets Y_t$
		\While{within the computational budget}
		\State $V^T \gets \Call{Search}{V^0}$
		\State Get~$\bm{\Gamma^{T}}$ from~$V^T$
		\State Adjust~$\bm{\rho^{+}}$ and~$\bm{\rho^{-}}$ according to~$\bm{\Gamma^{T}}$
		\State $S^{\text{tmp}}_t\gets\Call{Simulator} {S_t, \bm{\rho^{+}}, \bm{\rho^{-}}, D_t}$
		\State Update~$Y^{\text{tmp}}_t$ with~$S^{\text{tmp}}_t$
		\State $R \gets f_c(Y^{\text{tmp}}_t) - f_c(Z)$
		
		\State $\Call{Backpropagate}{V^T, R}$
		\If{ $R_{\text{top}} < R$}
		\State  $R_{\text{top}} \gets R$
		\State $Y_t \gets Y^{\text{tmp}}_t$
		\EndIf
		\EndWhile
		\EndWhile
		\EndFunction
	\end{algorithmic}
\end{algorithm}

\section{Experiments}\label{SecExp}
We first give the experimental setup including datasets, environmental model, and steganographic and steganalysis methods. Then we tune the hyperparameters of MCTSteg and compare MCTSteg and non-additive steganography in spatial and JPEG domains. Finally, we adopt ALASKA-v2 to compare MCTSteg and state-of-the-art machine learning-based steganographic methods, which gives a more comprehensive conclusion. The source codes and auxiliary materials are available for download from GitHub~\footnote{\url{https://github.com/tansq/MCTSteg}}.
\subsection{Experimental Setup}\label{SecExpSet}

\subsubsection{Dataset for the Environmental Model}

\emph{SZUBase} has 40,000 512$\times$512 full-resolution raw images. These images were collected by our laboratory, and converted with the same script of \emph{BOSSBase}~\cite{2011}. Due to copyright protection, they are not publicly available yet.

To protect the independence of training and testing datasets, prior research (including ASDL-GAN~\cite{2017Automatic}, UT-GAN~\cite{yang2019embedding}, and SPAR-RL~\cite{tang2020automatic}) utilizes \emph{SZUBase} to train the steganalyzer and evaluates the security performance with other datasets. Therefore, we also adopt \emph{SZUBase} for MCTSteg to train a SRNet~\cite{2018Deep} based environmental model. After scaling down to 256$\times$256 with “imresize” Matlab function, 38,000 cover-stego image pairs are used for training while the remaining 2,000 are used for validation. The stego images are generated by arbitrary steganography such as HILL~\cite{2015A} and JUNIWARD~\cite{Vojt2014Universal}. Specifically, in the JPEG domain, all images are decompressed without integer rounding. We train SRNet for 400k iterations with an initial learning rate of~$r_1 = 0.001$. Then the learning rate is decreased to~$r_2 = 0.0001$ for an additional 100k iterations, which is the same as the setting of Boroumand et al.\cite{2018Deep}. Finally, we select the trained model with the highest validation accuracy as our environmental model.

\subsubsection{Datasets for Performance Verification}
To compare MCTSteg with other state-of-the-art steganography, we use three datasets of \emph{BOSSBase} v1.10~\cite{2011}, \emph{BOWS2}~\cite{bows2.ec-lille.fr}, and \emph{ALASKA2}~\cite{alaska.utt.fr}. Both \emph{BOSSBase} and \emph{BOWS2} are composed of 10,000 grayscale images with size 512$\times$512 and ``pgm'' format. Due to the limited memory of our GPU (Tesla P100), we use ``imresize'' in Matlab with default settings to scale down those images to 256$\times$256. For the JPEG domain, we use ``imwrite'' in Matlab to transform the format from ``pgm'' to ``jpg'' with quality factors(QF) of 75 and 95. The \emph{ALASKA2} dataset contains 80,000 images of various sizes and formats. We use the 256$\times$256 size in both JPEG and spatial domains.

\subsubsection{Steganographic Method}

To prove that our MCTSteg can improve the security performance of additive steganography, we choose basic cost functions including S-(J)UNIWARD~\cite{Vojt2014Universal} and HILL~\cite{2015A}. Then we compare MCTSteg and four state-of-the-art non-additive steganographic frameworks, including CMD~\cite{2015CMD} and Synch~\cite{Denemark2015Improving} for the spatial domain, and BBC~\cite{li2018defining} and BBM~\cite{wang2020non} for the JPEG domain.

\subsubsection{Steganalyzer}

We use five state-of-the-art steganalyzers in JPEG and spatial domains for a more comprehensive comparison. They are based on either hand-crafted features or CNN structures. 
\begin{itemize}
	\item For hand-crafted feature-based steganalyzers, we adopt SRM~\cite{Fridrich2012Rich}, maxSRMd2~\cite{denemark2014selection}, and GFR~\cite{song2015steganalysis}. These steganalyzers are trained and tested on BOSSBase dataset, while the cover images and the corresponding stego images are pair-wisely and randomly split into a training set and a testing set with 1 : 1 proportion.
	\item For CNN structure-based steganalyzers, we adopt SRNet~\cite{2018Deep}, the most powerful one so far. Following the experimental settings and dataset split in ~\cite{2018Deep}, for QF75, SRNet is trained for 400k iterations with an initial learning rate of 0.001 and then the learning rate is decreased to 0.0001 for additional 100k iterations. For QF95, the SRNet models are finetuned from QF75 with the curriculum learning schedule in~\cite{2018Deep}.
	\item For computer vision models pretrained on ImageNet, we
          refine the EfficientNet B4 network\cite{yousfi2020imagenet,
            butora2021pretrain} on steganalysis tasks via curriculum
          learning\footnote{It is first refined on ALASKA2 dataset
            with the payload of 1.0 bpp/BPNZAC. Then it is finetuned
            via payload curriculum on BossBase+BOWS2 dataset.}. Please
          note that, for better detection performance, we have removed
          the stride from the first layer of EfficientNet B4 and the
          pair constrain of input data.

\end{itemize}

\subsection{Hyperparameter Tuning}\label{SecHyper}

\begin{figure}[!h]

	\includegraphics[scale=0.34]{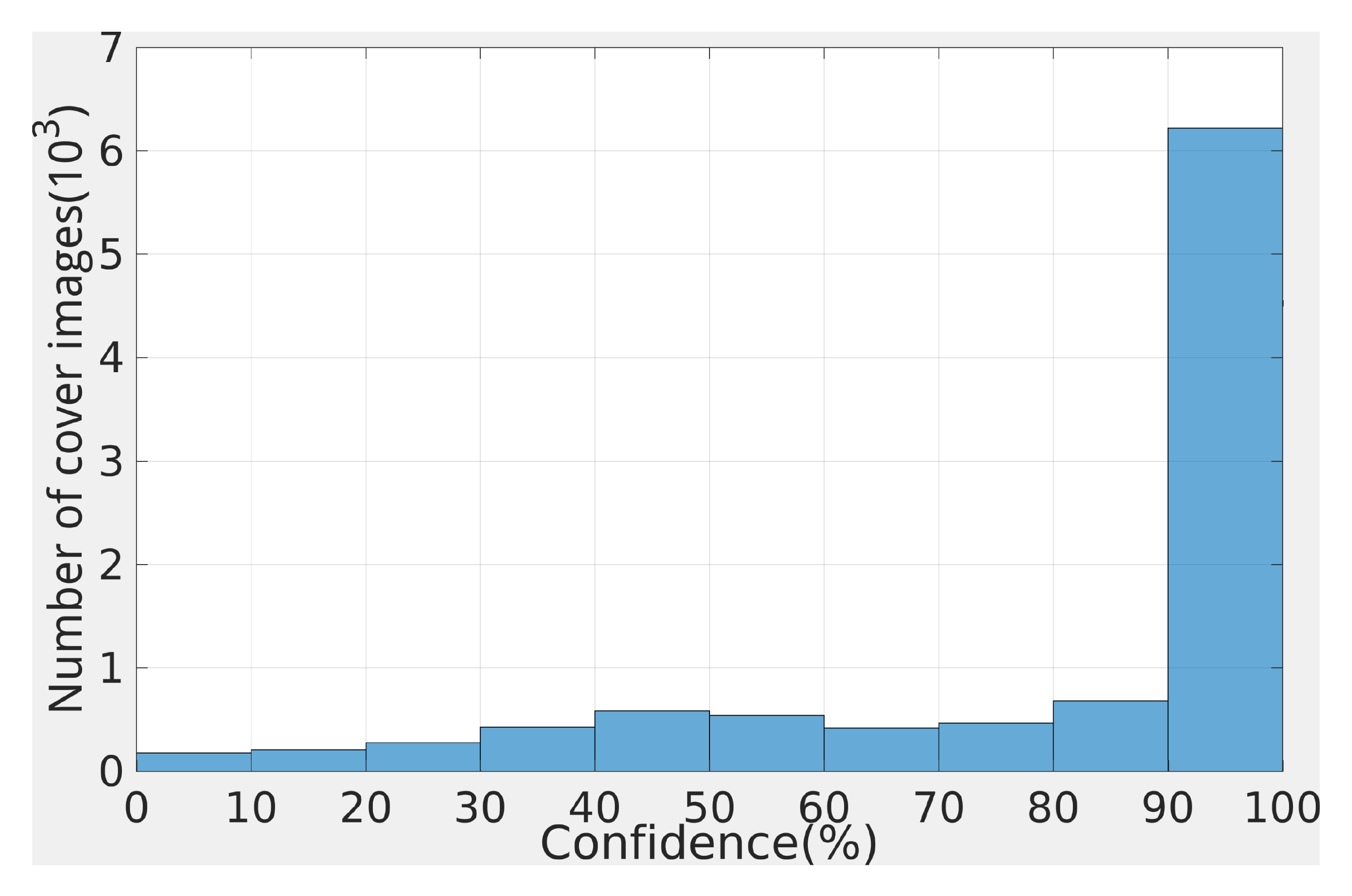}
	\caption{The confidence distribution of 10,000 cover images in BOSSBase.}
	
	\label{fig5}
	
\end{figure}

There are three major hyperparameters in MCTSteg, the maximum search count $M$, the confidence threshold and the scaling factor~$\alpha$ for distortion adjustment. $M$ balances the security performance and training time. It is empirically set to 128.

The proper confidence threshold can both ensure the sufficient security performance of final search result and avoid wasting superabundant time cost in search procedure. Therefore, we have conducted experiments on BOSSBase dataset to discover proper choice of the confidence threshold. In our experiment, the confidence represents for the percentage that environmental model classifies its input as ``cover''. And the result is showed in Fig. \ref{fig5}.

We can see that the confidence of most of the cover images are gathered in 90\%-100\%. Specifically, the confidence of 55.83\% cover images is higher than 98\%.
Thus, if the samples generated by MCTSteg can achieve the confidence higher than 98\%, the possibility that they are predicted by the environmental model as ``cover'' is comparable to more than half of the cover images in the dataset. As a consequence, we set the confidence threshold at 98\%.

\begin{figure}[ht]
	
	\centering
	\includegraphics[scale=0.70]{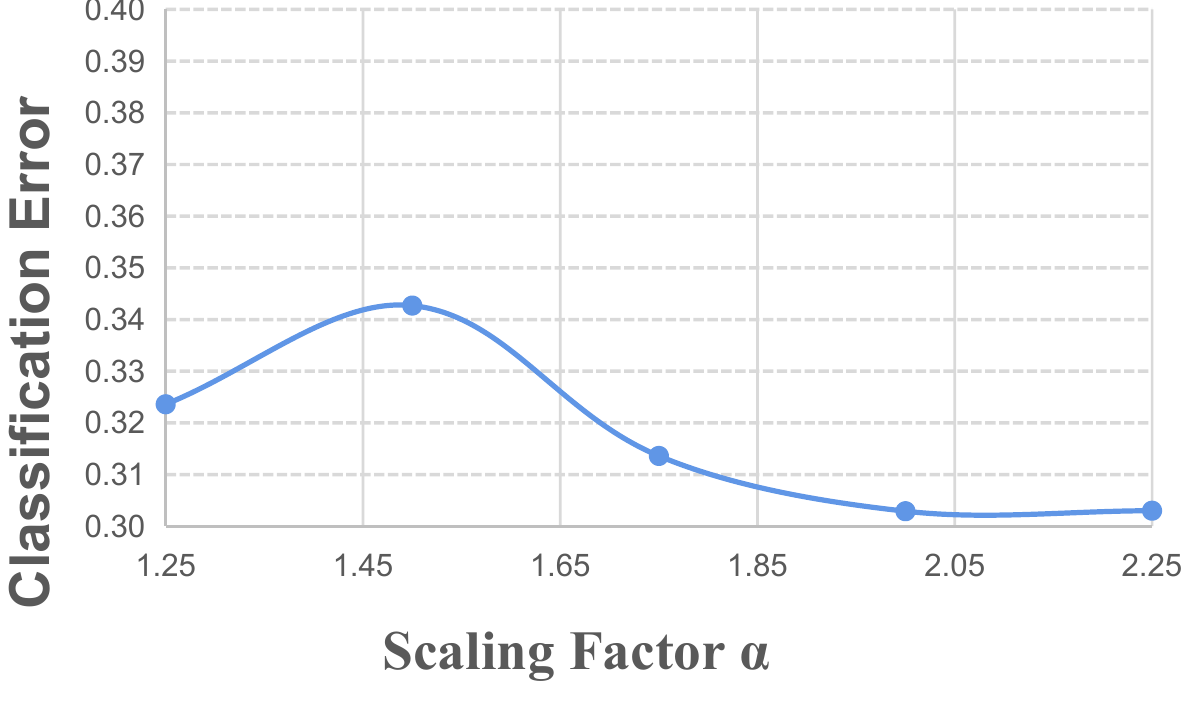}
	\caption{Classification error rates (maxSRMd2) of MCTSteg with different values of scaling factor~$\alpha$.\editorcomment{From the perspective of hyperparameter tuning, it is probably not enough to choose the optimal value of~$\alpha$ with only five data points.}}
	\label{fig8}
	
\end{figure}

The scaling factor~$\alpha$ adjusts distortion distribution as Equations~\ref{equ:omega_plus} and~\ref{equ:omega_minus} show.  With a payload of 0.4bpp, we use \emph{BOSSBase} and maxSRMd2 to search for the best value of~$\alpha$ in the spatial domain. From  Fig.\ref{fig8}, we find that the minimum classification error rate is achieved when~$\alpha = 1.5$. Therefore, for the rest of Sect. \ref{SecExp}, we adopt the~$M$ of $128$, the confidence threshold of 98\% and the~$\alpha$ of $1.5$.

\begin{figure*}[!ht]
	\centering
	\subfloat[05109.pgm]{
		\includegraphics[scale=0.30]{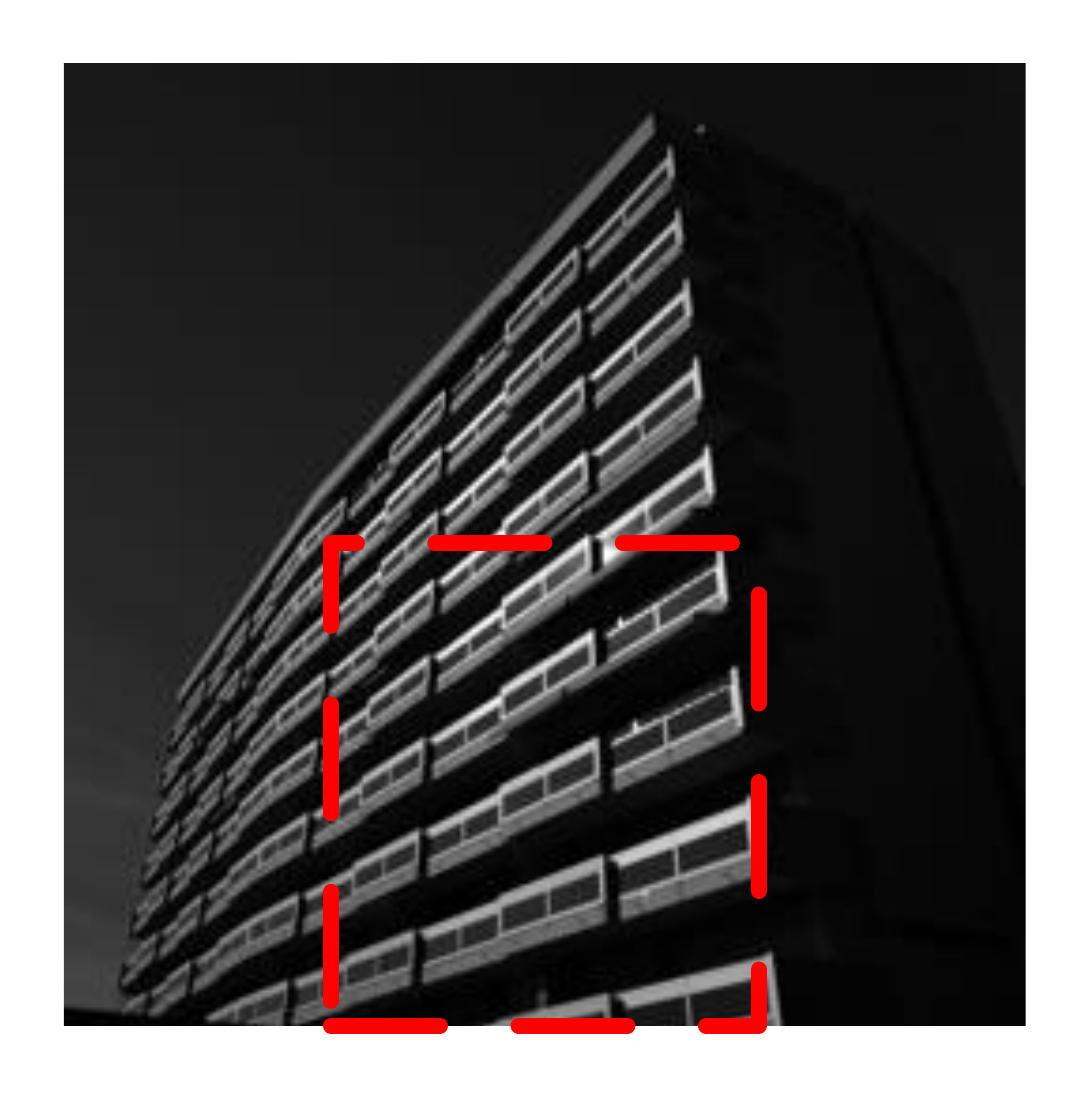}
	}
	\subfloat[SUNIWARD]{
		\includegraphics[scale=0.62]{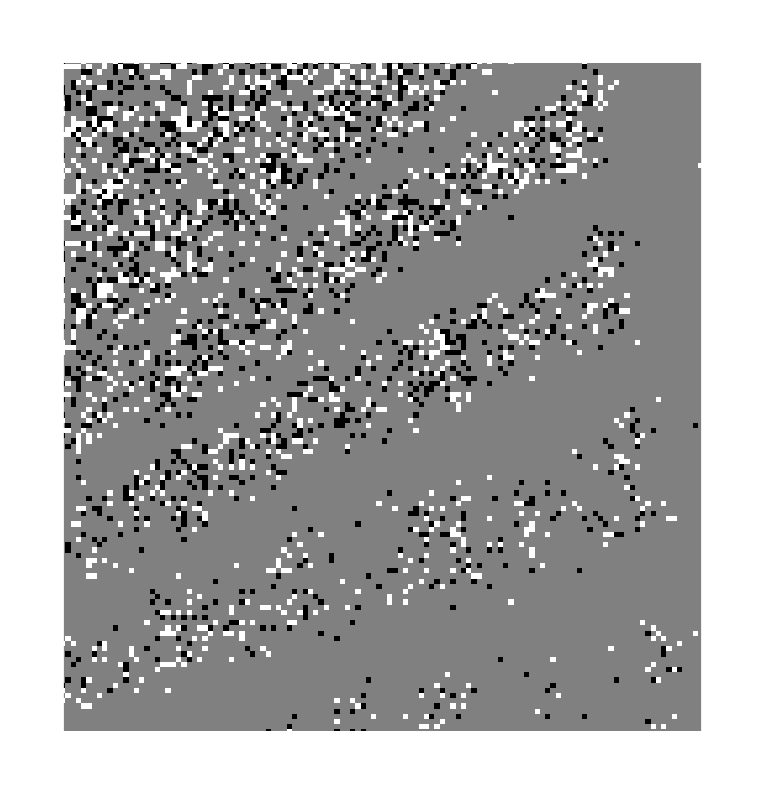}
	}
	\subfloat[SUNIWARD\_CMD]{
		\includegraphics[scale=0.62]{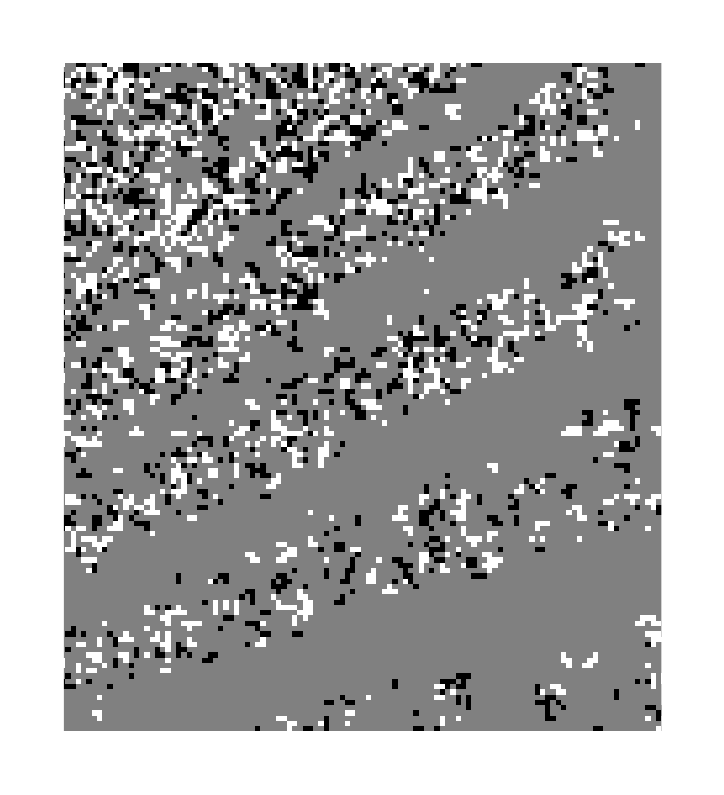}
	}
	\subfloat[SUNIWARD\_MCTS]{
		\includegraphics[scale=0.62]{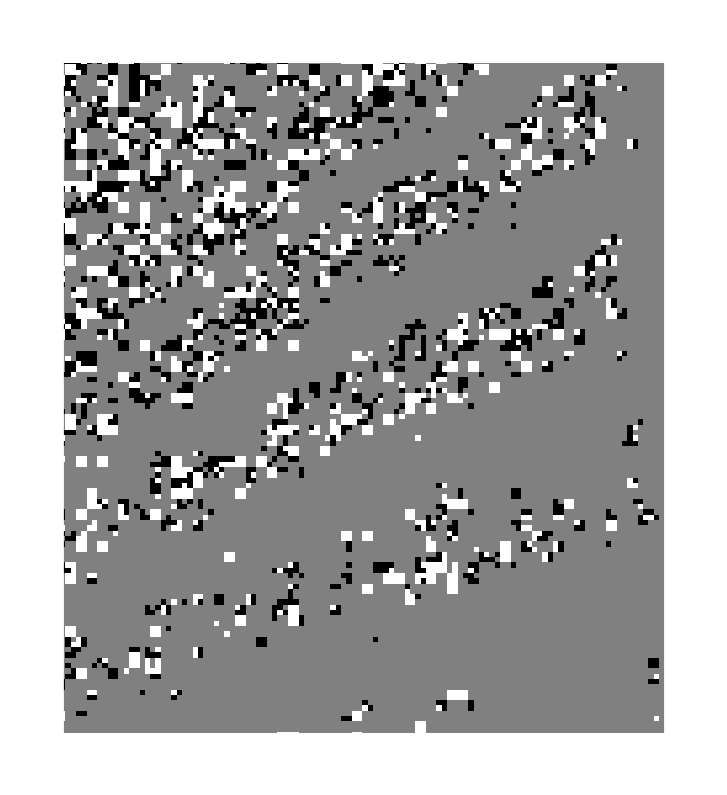}
	}
	\\
	\subfloat[Crop.pgm]{
		\includegraphics[scale=0.30]{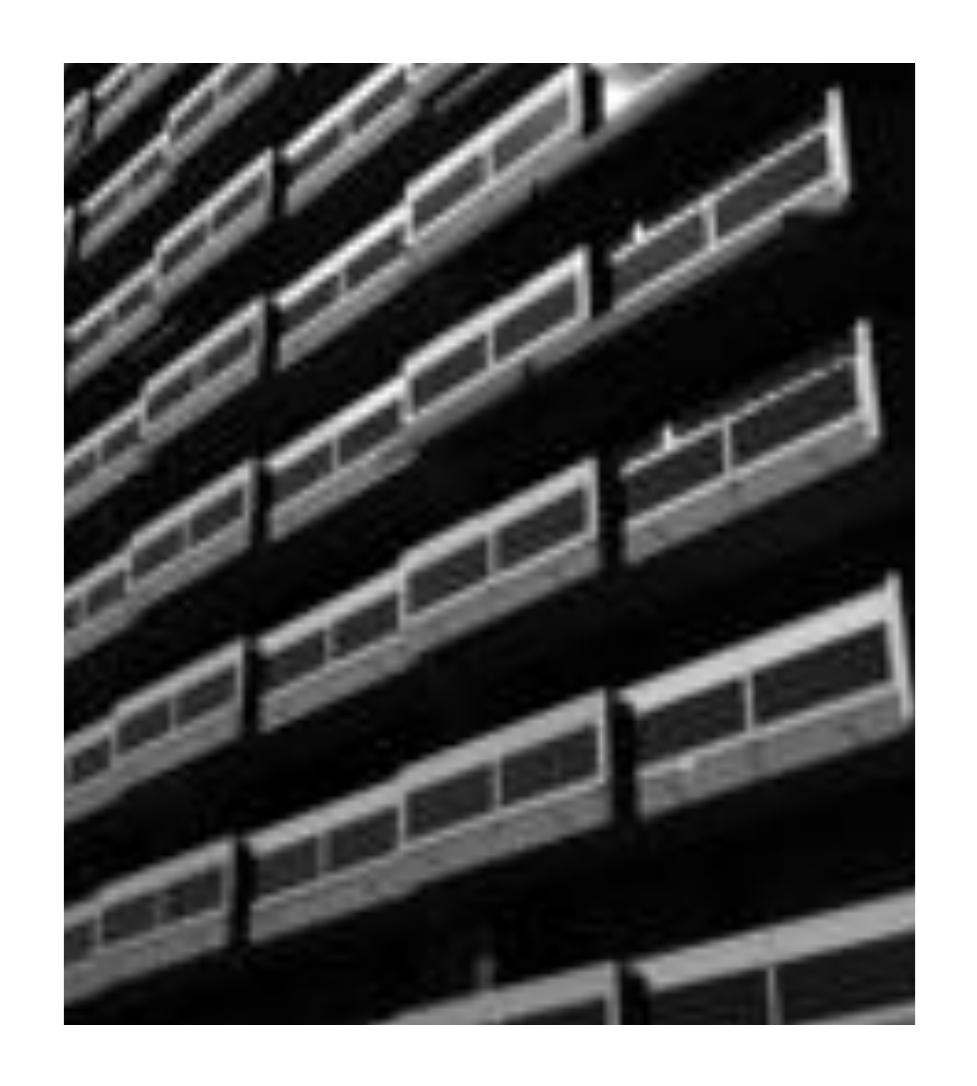}
	}
	\hspace{0.3cm}
	\subfloat[HILL]{
		\includegraphics[scale=0.62]{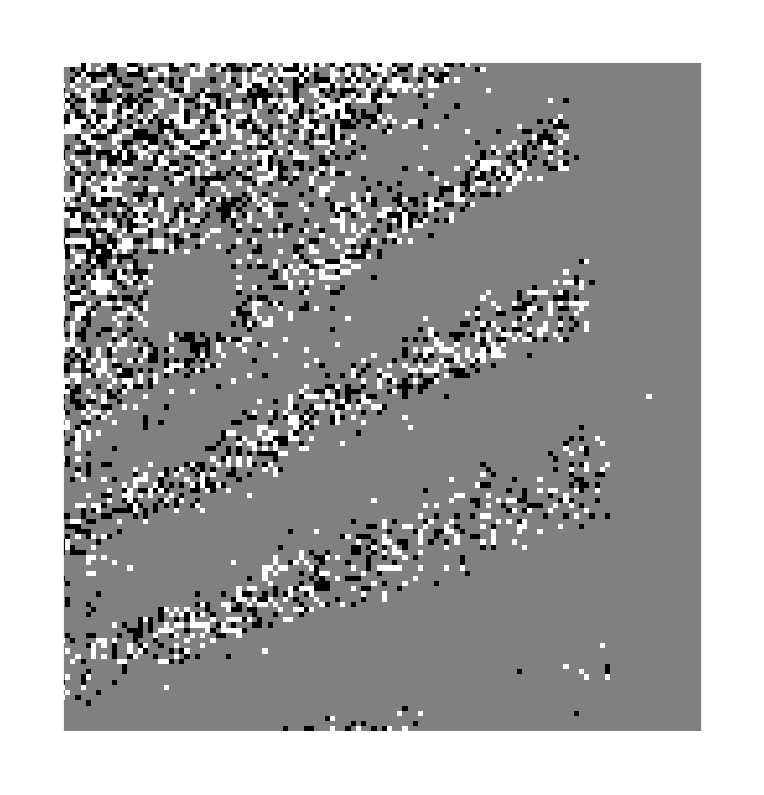}
	}
	\subfloat[HILL\_CMD]{
		\includegraphics[scale=0.62]{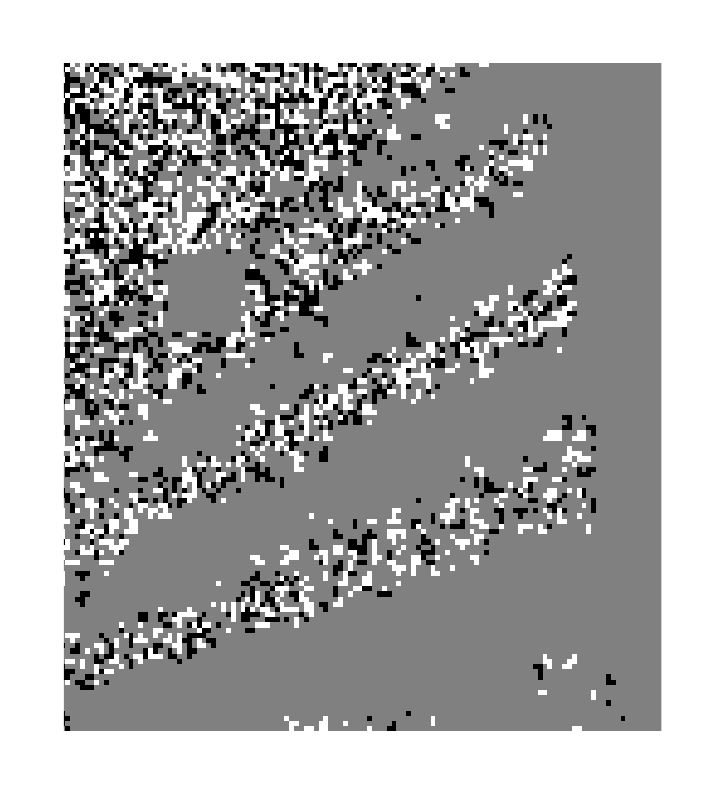}
	}
	\subfloat[HILL\_MCTS]{
		\includegraphics[scale=0.62]{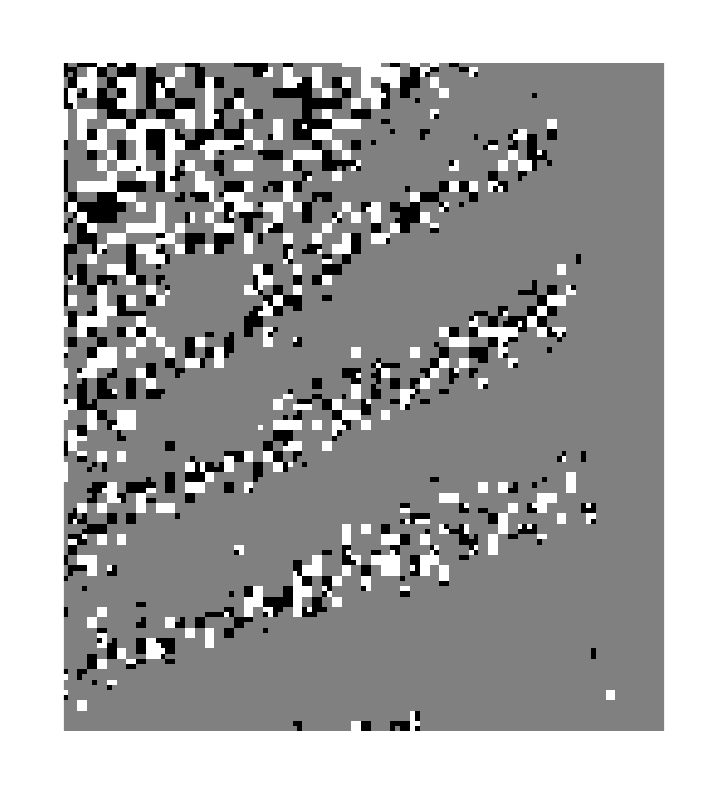}
	}
	\caption{(a) is a sample cover image where the area in the red rectangles is cropped into (b). (c) to (h) are the modification maps, where white pixels represent the modification of ``+1'' and dark pixels represent the modification of ``-1''.Due to possibly low printing resolution, readers are encouraged to zoom in the figures on a computer screen for better clarity.}
	
	\label{fig4}
\end{figure*}

\subsection{Visualizing Embedding Modification}\label{SecVis}
To make a visual comparison between the modification maps of different steganographic methods, we sample a cover image from BOSSBase v1.10 as Fig. \ref{fig4}(a) shows, which contains smooth regions, edges, and textured regions. Fig. \ref{fig4}(b) is cropped from Fig. \ref{fig4}(a) where the red rectangles denote the cropped area. For better visual effects, we embed the cover image with payload of 0.4 bits per pixel(bpp) by using additive cost functions including SUNIWARD, HILL and non-additive frameworks including CMD, MCTSteg. The modification maps are shown in Fig.~\ref{fig4}~(c) to (h). We can see that all steganographic methods prefer to embed message into the texture regions. However, the embedding modifications of MCTSteg are more concentrated into the center texture regions. On the contrary, for the boundaries between smooth regions and texture regions, there are fewer modifications of MCTSteg. Actually, manual strategies such as CMD will reduce the costs of pixels at those boundaries, which causes some pixels at the smooth regions are more likely to be modified. Moreover, the distributions of embedding modification between CMD and MCTSteg are obviously different, which means that the distortion adjustment strategies between MCTSteg and CMD are also different. Therefore, without human intervention, MCTSteg successfully develop its own novel strategies.

\subsection{Statistical Analysis}\label{SecSta}

\begin{table*}[!htp]
	\centering
	\caption{Statistical analysis of different steganographic methods on \emph{BOSSBase} with payload 0.4bpp}
	\label{Table0}
	\begin{tabular}{c|ccccc}
		\thickhline
		\cline{1-6}
		\textbf{Methods} & \textbf{$\alpha$} & \textbf{Change Rates} & \textbf{FCC (2nd Order)} & \textbf{FCC (3rd Order)} & \textbf{FCC (4th Order)} \\
		\thickhline
		\cline{1-6}
		HILL             & N/A               & 9.60\%                & 0.640\%                  & 0.113\%                  & 0.022\%                  \\
		\cline{1-6}
		HILL\_CMD        & 9                 & 12.4\%                & 1.805\%                  & 0.461\%                  & 0.101\%                  \\
		\cline{1-6}
		HILL\_MCTSteg    & \textbf{1.5}      & \textbf{9.06\%}       & \textbf{2.146\%}         & \textbf{0.509\%}         & \textbf{0.261\%}         \\
		\thickhline
		\cline{1-6}
	\end{tabular}
\end{table*}
\begin{table*}[!ht]
	\centering
	\caption{$P_E$ of different steganographic methods against SRM and maxSRMd2 in the spatial domain on \emph{BOSSBase}}
	\label{Table1}
	\begin{tabular}{|c|c|c|c|c|c|}
		\hline
		\multicolumn{2}{|c|}{\multirow{2}*{\textbf{Steganography}}} & \multicolumn{2}{|c|}{\textbf{SRM}} & \multicolumn{2}{|c|}{\textbf{maxSRMd2}}                                                                                                                      \\
		\cline{3-6}
		\multicolumn{2}{|c|}{}& \textbf{0.2bpp}& \textbf{0.4bpp}&\textbf{0.2bpp}&\textbf{0.4bpp}\\
		\hline
		\multirow{4}*{SUNIWARD}&Baseline&33.71\%&21.97\%&30.42\%& 20.49\%\\
		\cline{2-6}
		&Synch&39.98\%($\uparrow 6.27\%$)&30.14\%($\uparrow 8.17\%$)&40.26\%($\uparrow 9.84\%$)&30.22\%($\uparrow 9.73\%$)\\
		\cline{2-6}
		&CMD& 40.61\%($\uparrow 6.90\%$)& 30.58\%($\uparrow 8.61\%$)& 40.74\%($\uparrow 10.32\%$)&30.65\%($\uparrow 10.16\%$)\\
		\cline{2-6}
		& MCTSteg& 41.87\%($\bm{\uparrow} \textbf{8.16\%}$)& 34.42\%($\bm{\uparrow}$ \textbf{12.45\%}) &41.60\%($\bm{\uparrow}$ \textbf{11.18\%}) & 34.27\%($\bm{\uparrow}$ \textbf{13.78\%}) \\
		\hline
		\multirow{4}*{HILL}& Baseline& 38.40\%& 27.93\%&32.94\%& 23.88\%\\
		\cline{2-6}
		& Synch&43.11\%($\uparrow 4.71\%$)&35.29\%($\uparrow 7.36\%$)&43.08\%($\uparrow 10.14\%$)&35.23\%($\uparrow 11.35\%$)\\
		\cline{2-6}
		& CMD& 43.51\%($\uparrow 5.11\%$)& 35.95\%($\uparrow 8.02\%$)& 43.34\%($\uparrow 10.40\%$)& 35.86\%($\uparrow 11.98\%$)\\
		\cline{2-6}	& MCTSteg & 44.01\%($\bm{\uparrow}$ \textbf{5.61\%})& 39.33\%($\bm{\uparrow}$ \textbf{11.40\%}) & 43.67\%($\bm{\uparrow}$ \textbf{10.73\%}) & 39.17\%($\bm{\uparrow}$ \textbf{15.29\%}) \\
		\hline
	\end{tabular}
\end{table*}

\begin{table*}[!ht]
	\centering
	\caption{$P_E$ of different steganographic methods against SRNet on \emph{BOSSBase}+\emph{BOWS2} in the spatial domain}
	\label{Table2}
	\begin{tabular}{c|ccccc}
		\thickhline
		\cline{1-6}
		\textbf{Basic additive distortion}&\textbf{Payload} & \textbf{Baseline} & \textbf{Synch} & \textbf{CMD} & \textbf{MCTSteg} \\
		
		\thickhline
		\cline{1-6}

		\multirow{2}*{S-UNIWARD}&0.2bpp & 21.16\% &28.31\%($\uparrow 7.15\%$)& 28.45\%($\uparrow 7.29\%$) & \textbf{29.81}\%($ {\uparrow}\textbf{8.65\%}$)  \\
		\cdashline{2-6}
		&0.4bpp &12.56\% &16.20\%($\uparrow 3.64\%$)& 16.23\%($\uparrow 3.67\%$) & \textbf{16.57}\%($ {\uparrow} $\textbf{4.01\%})\\
		\thickhline 
		\cline{1-6}
		\multirow{2}*{HILL}&0.2bpp & 24.74\% &32.17\%($\uparrow 7.43\%$)&32.29\%($\uparrow 7.55\%$) & \textbf{33.32}\%($ {\uparrow} \textbf{8.58\%}$) \\

		\cdashline{2-6}
		& 0.4bpp&16.14\% &21.93\%($\uparrow 5.79\%$)& 22.14\%($\uparrow 6.00\%$) & \textbf{22.74}\%($ {\uparrow} \textbf{6.60\%}$) \\
		\thickhline 
		\cline{1-6}
	\end{tabular}
\end{table*}

To analyze statistically, we adopt change rates and FCC~\cite{2015CMD} as evaluation metrics. In \cite{2015CMD}, the authors figured out that clustering modification directions can help improve security performance of steganography in spatial domain. Thus, they proposed the n-th order FCC metric to compute the average frequency of occurrences in the row/column direction for consecutive positive/negative modification polarity. The~$n$-th order FCC, denoted as~$F(n)$, is defined by:
\begin{equation}
	F(n) = \frac{1}{4}(H(n,1)+H(n,-1)+V(n,1)+V(n,-1)),
\end{equation}
where
\begin{eqnarray}
	H(n,k) =& \frac{\sum_{i=1}^{n_1}\sum_{j=1}^{n_2-n+1}(\delta(d_{i,j}-k)\ldots\delta(d_{i,j+n}-k))}{n_1(n_2-n+1)}\\
	V(n,k) =& \frac{\sum_{i=1}^{n_1-n+1}\sum_{j=1}^{n_2}(\delta(d_{i,j}-k)\ldots\delta(d_{i+n,j}-k))}{(n_1-n+1)n_2},
\end{eqnarray}
where~$\delta(\cdot)$ is an impulse function and~$d$ is a modification map. Generally speaking, the higher the n-th FCC score, the stronger effect of the clustering modification directions. Therefore, the n-th FCC is a relevant and low-cost metric for evaluating the security performance of non-additive steganography.

The FCC scores and change rates are shown in Tab.~\ref{Table0}. We can see that MCTSteg's second- to fourth-order FCC scores are much higher than those of other methods. For example, MCTSteg's fourth-order FCC score is even more than twice of CMD's. It means that without the guidance of human, MCTSteg has automatically learned the strategy of clustering modification directions from the interaction between MCTS and environmental model.

For the change rate, we can see that it is 9.60\% for HILL and 12.40\% for CMD HILL, which means that the security performance improvement of CMD comes with additional increment in change rate. But, for MCTSteg HILL, its change rate is 9.06\%, which is even smaller than HILL. Therefore, the security performance improvement of MCTSteg does not cause an additional cost of change rate.

\subsection{Comparison of state-of-the-art Non-additive Methods}\label{SecNon}

To compare security performance of different steganographic methods, we adopt detection error rate $P_E$ on testing set, which is calculated by the false alarm rate $P_{FA}$ and the missed detection rate $P_{MD}$ as follows:
\begin{equation}
	P_{E} = \mathop{min}\limits_{P_{FA}}\frac{1}{2}(P_{FA}+P_{MD}).
\end{equation}

In the spatial domain, we adopt SUNIWARD and HILL as basic additive cost functions and compare security performance between Synch, CMD, and MCTSteg with payloads of 0.2 and 0.4bpp. Detected by SRM and maxSRMd2, the results on \emph{BOSSBase} are shown in Tab.~\ref{Table1}.

From Tab.~\ref{Table1}, we can see that MCTSteg achieves substantial improvements based on both SUNIWARD and HILL. Under the payload of 0.4bpp and detected by maxSRMd2, the improvement of our MCTSteg\_HILL is \textbf{15.29\%}, which is \textbf{3.31\%} higher than that of CMD\_HILL. As for SRM, the performance margin between MCTSteg and other enhancement methods is also significant.

From Tab.~\ref{Table2}, as we can see, all of these non-additive steganographic frameworks increase the detection error rates against SRNet. But, MCTSteg achieves the highest improvements in all scenarios, for example, the improvement of MCTSteg over the original S-UNIWARD at the payload of 0.2bpp is \textbf{8.65\%} while CMD, the second place, only achieves the improvement of \textbf{7.29\%}.

\begin{table*}[!htbp]
	
	\centering\caption{$P_E$ of different steganographic methods against EfficientNet B4 on \emph{BOSSBase+BOWS2} in the spatial domain}
	\label{Table3}
	\begin{tabular}{c|cccc}
		\thickhline
		\cline{1-5}
		
		\textbf{Basic additive distortion} & \textbf{Payload}&\textbf{Baseline}  & \textbf{CMD} & \textbf{MCTSteg} \\
		\thickhline
		\cline{1-5}

		\multirow{2}*{S-UNIWARD} &0.2bpp& 10.53\%  & 13.72\% $(\uparrow 3.19\%)$& \textbf{33.70\%} $(\uparrow \textbf{23.17\%})$\\
		\cdashline{2-5}
		&0.4bpp &4.08\% &  6.07\% $(\uparrow 1.99\%)$& \textbf{23.06\%} $(\uparrow \textbf{18.98\%})$ \\
		
		\thickhline
		\cline{1-5}
		
		\multirow{2}*{HILL} &0.2bpp & 10.23\% & 14.32\% $(\uparrow 4.09\%)$& \textbf{38.72\% }$(\uparrow \textbf{28.49\%})$\\
		\cdashline{2-5}
		& 0.4bpp& 4.96\%  & 6.42\% $(\uparrow 1.46\%)$& \textbf{29.40\%} $(\uparrow \textbf{24.44\%})$\\
		\thickhline
		\cline{1-5}
	\end{tabular}
	
\end{table*}

\begin{table*}[!h]
	\centering
	
	\caption{$P_E$ of different steganographic methods based on J-UNIWARD against GFR on \emph{BOSSBase} in the JPEG domain with QF=75 and 95}
	\label{Table4}
	\begin{tabular}{c|c|cccc}
		\thickhline
		\cline{1-6}
		\textbf{QF} &\textbf{Payload}& \textbf{Baseline}& \textbf{BBM}& \textbf{BBC-BBM}& \textbf{MCTSteg} \\
		\thickhline
		\cline{1-6}
		\multirow{2}*{75}&0.2bpnzac& 39.71\% & 40.62\%($\uparrow 0.91\%$) & 40.78\%($\uparrow 1.07\%$) & \textbf{40.84\%}(${\uparrow} \textbf{1.13\%}$) \\
		\cdashline{2-6}
		&0.4bpnzac& 25.39\% & 27.91\%($\uparrow 2.52\%$) & 28.71\%($\uparrow 3.32\%$) & \textbf{29.12\%}(${\uparrow} \textbf{3.70\%}$) \\
		\thickhline
		\cline{1-6}
		\multirow{2}*{95}&0.2bpnzac& 46.06\%  & 46.50\% $(\uparrow 0.44\%)$ & 46.34\% $(\uparrow 0.28\%)$& \textbf{46.89\%}$(\uparrow \textbf{0.83\%})$\\
		\cdashline{2-6}
		& 0.4bpnzac & 38.15\% & 39.43\% $(\uparrow 1.28\%)$& 39.56\%$(\uparrow 1.41\%)$ & \textbf{39.98\%} $(\uparrow \textbf{1.83\%})$\\
		\thickhline
		\cline{1-5}
	\end{tabular}
\end{table*}

\begin{table*}[!h]
	\centering
	\caption{$P_E$ of different steganographic methods based on J-UNIWARD against SRNet on \emph{BOSSBase}+\emph{BOWS2} in the JPEG domain with QF=75 and 95}
	
	\label{Table5}
	\begin{tabular}{c|c|cccc}
		\thickhline
		\cline{1-6}
		\textbf{QF} &\textbf{Payload}& \textbf{Baseline} & \textbf{BBM} & \textbf{BBC-BBM} & \textbf{MCTSteg} \\
		\thickhline
		\cline{1-6}
		\multirow{2}*{75}&0.2bpnzac& 19.40\%  & 23.50\%($\uparrow 4.10\%$) & 24.12\%($\uparrow 4.72\%$) & \textbf{26.67\%}($ {\uparrow} \textbf{7.27\%}$) \\
		\cdashline{3-6}
		&0.4bpnzac& 7.73\% & 8.37\%($\uparrow 0.64\%$) & 9.10\%($\uparrow 1.37\%$) &\textbf{ 13.47\%}($ {\uparrow} \textbf{5.74\%}$) \\
		\thickhline
		\cline{1-6}
		\multirow{2}*{95}&0.2bpnzac& 36.15\%  & 37.86\% $(\uparrow 1.71\%)$ & 38.43\%($\uparrow 2.28\%$) & \textbf{39.91\%}($\uparrow \textbf{3.76\%}$) \\
		\cdashline{3-6}
		&0.4bpnzac& 19.68\% & 22.43\%$(\uparrow 2.75\%)$ & 22.26\%$(\uparrow 2.58\%)$  & \textbf{26.91\%}$(\uparrow \textbf{7.23\%})$ \\
		\thickhline
		\cline{1-6}
	\end{tabular}
	
\end{table*}

\begin{table*}[!h]
	\centering
	\caption{$P_E$ of different steganographic methods based on J-UNIWARD against EfficientNet B4 on \emph{BOSSBase}+\emph{BOWS2} in the JPEG domain with QF=75}
	\label{Table6}
	\begin{tabular}{c|cccc} 
		\thickhline
		\cline{1-5}
		\textbf{Payload} & \textbf{Baseline} & \textbf{BBM} & \textbf{BBC-BBM} & \textbf{MCTSteg} \\
		\thickhline
		\cline{1-5}
		0.2bpnzac & 27.40\% & 28.38\% $(\uparrow 0.98\%)$ & 28.10\%  $(\uparrow 0.70\%)$ & \textbf{32.05\%}  $(\uparrow \textbf{4.65\%})$ \\
		
		\cdashline{2-5}
		0.4bpnzac & 12.66\% & 14.73\%  $(\uparrow 2.07\%)$ &14.33\%  $(\uparrow 1.67\%)$ & \textbf{16.07\%}  $(\uparrow \textbf{3.41\%})$ \\
		
		\thickhline
		\cline{1-5}
	\end{tabular}
\end{table*}

From Tab.~\ref{Table3}, we can see that, under the detection of ImageNet pretrained EfficientNet B4, the improvements of MCTSteg over CMD are significant for both basic additive distortions. For instance, for the payload of 0.4 bpp, on top of S-UNIWARD, the $P_E$ of MCTSteg is \textbf{23.06\%} while CMD, the second place, is 6.07\%. Specially, no matter which basic additive distortion is adopted, the initial network weights of EfficientNet B4 are identical to our proposed MCTSteg and other schemes at the payload of 0.4 bpp,

In the JPEG domain, we adopt JUNIWARD as basic additive steganography and compare security performance among MCTSteg, BBM~\cite{wang2020non}, and BBC-BBM (i.e. the combination of BBC~\cite{li2018defining} and BBM) with QF=75 and 95, because Wang et al.~\cite{wang2020non} report that BBC-BBM achieves better security performance than either BBC or BBM.

\begin{table*}[!ht]
	
	\centering
	\caption{$P_E$ of different steganographic methods against SRM and maxSRMd2 on ALASKA-v2 in the spatial domain}
	\label{Table7}
	\renewcommand\arraystretch{1.5}
	\begin{tabular}{cccc}
		\thickhline
		\cline{1-4}
		\textbf{Steganalyzer} & \textbf{Steganographic Method} & \textbf{0.2 bpp}  & \textbf{0.4 bpp} \\ \thickhline
		\cline{1-4}
		\multirow{6}*{SRM} & HILL & 45.04\%  & 39.23\% \\
		
		\cdashline{2-4} & ASDL-GAN  & 41.24\% $(\downarrow 3.80\%)$ & 36.95\% $(\downarrow 2.28\%)$\\
		\cdashline{2-4} & UT-GAN & 44.87\% $(\downarrow 0.17\%)$ & 39.31\% $(\uparrow 0.08\%)$\\
		\cdashline{2-4} & SPAR-RL & 45.13\% $(\uparrow 0.09\%)$& 40.63\% $(\uparrow 1.40\%)$\\
		\cdashline{2-4} & MCTSteg\_HILL & 47.61\%$\bm{\uparrow} (\textbf{2.57\%})$ & 44.57\%$\bm{\uparrow} (\textbf{5.34\%})$ \\
		\cline{1-4}
		\multirow{6}*{maxSRMd2} & HILL & 44.53\% & 38.55\% \\
		
		\cdashline{2-4} & ASDL-GAN & 41.85\% $(\downarrow 2.68\%)$& 36.83\% $(\downarrow 1.72\%)$\\
		\cdashline{2-4} & UT-GAN & 44.76\% $(\uparrow 0.23\%)$& 40.29\% ($\uparrow 1.74\%$)\\
		\cdashline{2-4} & SPAR-RL & 44.28\% $(\downarrow 0.25\%)$& 41.36\% $(\uparrow 2.81\%)$\\
		\cdashline{2-4} & MCTSteg\_HILL & 47.89\%($\bm{\uparrow}$ \textbf{3.34\%}) & 44.97\%($\bm{\uparrow} \textbf{6.42\%}$) \\\thickhline
		
		\cline{1-4}
	\end{tabular}
\end{table*}

\begin{table*}[!htbp]
	
	\centering
	\caption{$P_E$ of different steganographic methods with STC on \emph{BOSSBase} in the spatial domain}
	\label{Table8}
	\renewcommand\arraystretch{1.5}
	\begin{tabular}{ccccc}
		\thickhline
		\cline{1-5}
		\textbf{Steganalyzer} & \textbf{Schemes} & \textbf{embedder} &\textbf{0.2 bpp}  & \textbf{0.4 bpp} \\ 
		\thickhline
		\cline{1-5}
		\multirow{10}*{SRM} & \multirow{2}*{HILL} & Simulator & 38.40\%  & 27.93\%  \\
		\cdashline{3-5}
		&& STC &37.51\% $(\downarrow 0.89\%)$ & 27.39\% $(\downarrow 0.54\%)$ \\
		\cdashline{2-5} 
		& \multirow{2}*{UT-GAN} & Simulator & 38.26\% & 29.14 \% \\
		\cdashline{3-5}
		&& STC &37.68\% $(\downarrow 0.58\%)$ & 27.90 \%  $(\downarrow 1.24\%)$ \\
		\cdashline{2-5} 
		& \multirow{2}*{SPAR-RL}  & Simulator & 39.17\% & 29.15\% \\
		\cdashline{3-5}
		&& STC &38.29\% $(\downarrow 0.88\%)$& 28.22\% $(\downarrow 0.93\%)$ \\
		\cdashline{2-5} 
		& \multirow{2}*{MCTSteg\_HILL} & Simulator & 44.01\% & 39.33\% \\
		\cdashline{3-5}
		&& STC &\textbf{40.46\%} $(\downarrow \textbf{3.55\%})$& \textbf{30.12\%} $(\downarrow \textbf{9.21\%})$ \\
		\cdashline{2-5} 
		& \multirow{2}*{CMD\_HILL} & Simulator &  43.51\% & 35.95\% \\
		\cdashline{3-5}
		&& STC &43.08\% $(\downarrow 0.43\%)$& 34.14\% $(\downarrow 1.81\%)$ \\
		
		\thickhline
		\cline{1-5}
	\end{tabular}
\end{table*}

For traditional hand-crafted feature-based steganalyzers, we select GFR and its results are shown in Tab.~\ref{Table4}. It can be seen that MCTSteg achieves highest security performance among the state-of-the-art steganographic schemes in JPEG domain. For example, with the payload of 0.4BPNZAC and QF75, MCTSteg achieves the highest security performance which is \textbf{3.70\%} higher than JUNIWARD. For QF95, the most secure scheme is still MCTSteg.

From Tab.~\ref{Table5}, we can see that, adopted SRNet as steganalyzer, the gaps in $P_E$ between MCTSteg and other methods are substantial at both QF75 and QF95. For example, with QF=75, MCTSteg's improvement is \textbf{7.27\%} while BBC-BBM, the second place, achieves the improvement of \textbf{4.72\%} with a payload of 0.2 BPNZAC. For QF95, the improvements of MCTSteg over BBM or BBC-BBM are more obvious. For instance, with payload of 0.4 BPNZAC, MCTSteg can increase $P_E$ of SRNet by 7.23\% while BBM, the second place, can only increase $P_E$ by 2.75\%.

From Tab.~\ref{Table6}, we can see that MCTSteg has achieved the best security performance against ImageNet pretrained EfficientNet B4 model with QF75. For example, with the payload of 0.2 BPNZAC, MCTSteg's improvement is \textbf{4.65\%} while BBM, the second place, is $0.98\%$.

Overall, our proposed MCTSteg achieves the best security performance against both hand-crafted feature-based and deep learning-based steganalyzers. Specially,  experimental results of a more complex EfficientNet B4 model have also demonstrated that, with SRNet as the environmental model, MCTSteg can still achieve the remarkable security performance improvement over CMD in spatial domain, or BBM/BBC-BBM in JPEG domain. To some extent we can make a conclusion that MCTSteg has effectively learned the ``adversarial component'' between steganography and steganalysis in both spatial and JPEG domain. However, we have to acknowledge that we have no explanation why the security performance of MCTSteg is inconsistent among different steganalysis scenarios. After all, interpretability of deep-learning based solutions is still a tough challenge to the researchers. We will devote ourselves to this issue in our future research.

\subsection{Comparison with Machine Learning-Based Methods}\label{SecSOTA}
In this section, we adopt ASDL-GAN~\cite{7350138}, UT-GAN~\cite{yang2019embedding}, and SPAR-RL~\cite{tang2020automatic} for comparison, while HILL is the baseline. Considering the time-cost problem, we use SRM and maxSRMd2 to evaluate the security performance of these methods with payloads of 0.2 and 0.4bpp. The ALASKA-v2 dataset is split into a training set and a test set of the same size (40,000 either). The results are shown in Tab.~\ref{Table5}.

We can see that under the payload of 0.4bpp and detected by SRM and maxSRMd2, the security performance of our method is \textbf{5.34\%} and \textbf{6.42\%} higher than HILL respectively. For other methods, ASDL-GAN is obviously worse than HILL, UT-GAN and SPAR-RL are better than HILL but worse than MCTSteg. \par
As for the payload of 0.2bpp, MCTSteg achieves \textbf{47.61\%} and \textbf{47.89\%} error rate, which means that the classification result of SRM and maxSRMd2 is close to random prediction. Meanwhile the security performance of ASDL-GAN is lower than HILL while UT-GAN and SPAR-RL achieve comparable security performance of HILL. Therefore, from Tab.~\ref{Table5}, we get a conclusion that MCTSteg achieves the best security performance among all the machine learning-based steganographic methods.

\subsection{Security Performance of Embedding with STC}\label{STC}
In this section, we use STC to embed secret message instead of using optimal embedding simulator. Then we conduct experiments against SRM on BOSSBase dataset in the spatial domain to compare security performance of our proposed MCTSteg and other state-of-the-art steganographic schemes. The results are shown in Tab.~\ref{Table7}.

\begin{figure}[!h]
	\centering
	\vspace{-0.7cm}
	\subfloat[01013.pgm]{\label{fig12a}
		\includegraphics[scale=0.25]{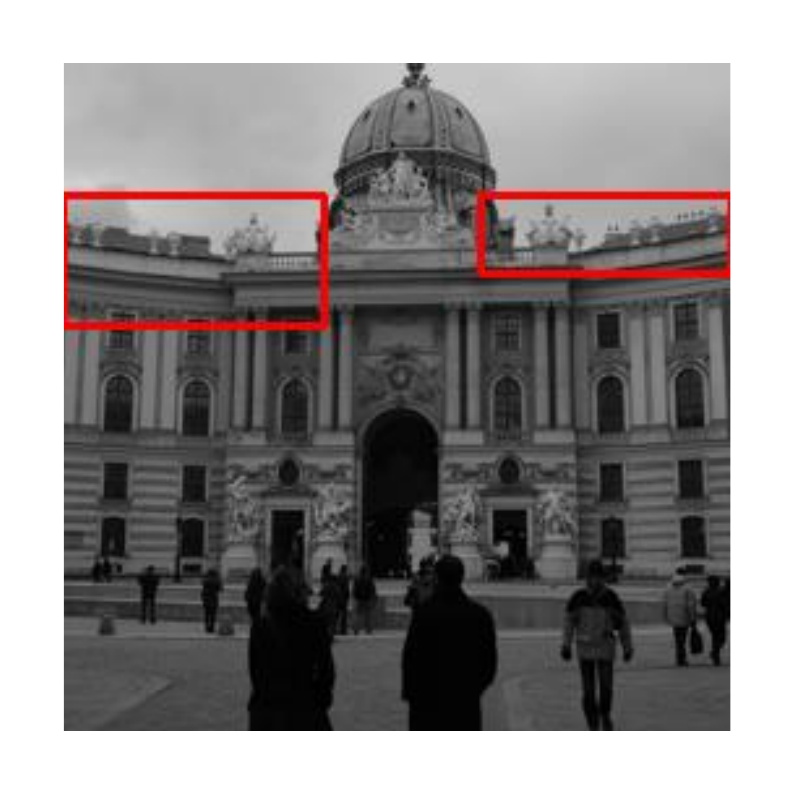}}
	\\
	\vspace{-1cm}
	\subfloat[MCTSteg\_HILL\_Simulator\newline Left Box]{\label{fig7b}
		\includegraphics[width=0.5\linewidth]{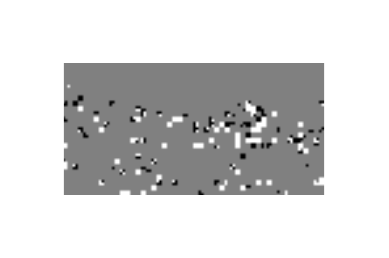}}
	\hspace{-1cm}
	\subfloat[MCTSteg\_HILL\_STC\newline Left Box]{\label{fig7c}
		\includegraphics[width=0.5\linewidth]{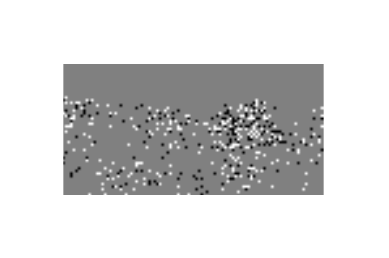}}
	\\
	\vspace{-1cm}
	\subfloat[MCTSteg\_HILL\_Simulator\newline Right Box]{\label{fig7d}
		\includegraphics[width=0.5\linewidth]{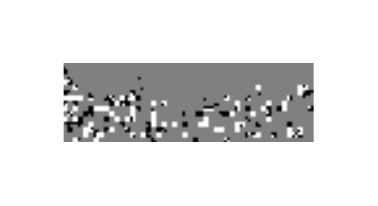}}
	\hspace{-1cm}
	\subfloat[MCTSteg\_HILL\_STC\newline Right Box]{\label{fig7e}
		\includegraphics[width=0.5\linewidth]{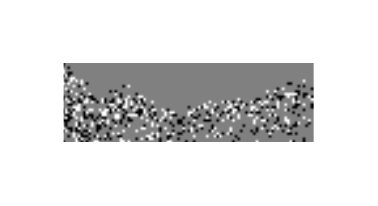}}
	\caption{(a) is the 01013.pgm from BOSSBase. Note that the red rectangles of (a) are the regions with the most obvious difference of modification maps between different embedders. (b)-(e) are the corresponding modification maps cropped by the red rectangles, which are embedded by MCTSteg\_HILL~with different embedders respectively.}
	\label{fig7}
	
\end{figure}

From Tab. \ref{Table7}, we can see that with STC, though MCTSteg\_HILL achieves the best security performance among the state-of-the-art deep-learning based steganographic schemes, i.e. ASDL-GAN, UT-GAN and SPAR-RL, its security performance is inferior to CMD\_HILL, which is different from the case of using optimal embedding simulator.
From Tab. \ref{Table7}, it can be found that for MCTSteg\_HILL, the security performance gap between the one with STC and the one with optimal embedding simulator is pronouncedly larger than other mentioned schemes. To further investigate the problem behind it, we compare the modification maps embedded by STC and the corresponding optimal embedding simulator, as shown in Fig.~\ref{fig7}.

From Fig. \ref{fig7}, it can be easily found that the distributions of modifications direction and location are completely different between Fig. \subref{fig7b} and Fig. \subref{fig7c} as well as Fig. \subref{fig7d} and Fig. \subref{fig7e}. To some extent, compared with the case of using optimal embedding simulator, the block effect of both direction and location of modifications has faded in Fig. \subref{fig7c} and Fig. \subref{fig7c}. As a result, they are more similar to modification maps of additive cost function. 

Since the optimal embedding simulator represents for the theoretical performance upper bound while the STC is trying to approach this bound. Therefore, it is possible for STC to cause a large coding loss in some steganographic schemes such as MCTSteg. As a result, for MCTSteg, this coding loss causes the pronouncedly giant security performane gap in Tab. \ref{Table7} and obviously visual difference of modification maps in Fig. \ref{fig7}.

\subsection{Computational Cost}\label{CC}

\begin{table*}[!h]
	\centering
	\caption{Comparison of the Number of Parameters and Computational Cost for Different Steganographic Methods }
	
	\label{Table9}
	\renewcommand\arraystretch{1.5}
	\begin{threeparttable}[!ht]
		\begin{tabular}{cccccc} 
			
			\cline{1-6}
			\thickhline
			\textbf{Models} & \textbf{Metrics} & \textbf{ASDL-GAN} & \textbf{UT-GAN} & \textbf{SPAR-RL} & \textbf{MCTSteg}\\
			\thickhline
			\cline{1-6}
			\textbf{Generator /} & Parameters & 1.63 $\times 10^5$  & $2.59 \times 10^6$ & $2.59 \times 10^6$ & -\\
			\cdashline{2-6}
			\textbf{Policy network} & FLOPs & $1.07 \times 10^{10}$ & $2.70 \times 10^8$ & $2.70 \times 10^8$ & -\\
			\cline{1-6}
			\textbf{Embedding} & Parameters & $3 \times 10^2$ & - & - & - \\
			\cdashline{2-6}
			\textbf{Simulator} & FLOPs & $1.97 \times 10^7$ & $6.55 \times 10^4 $ & $6.55 \times 10^4 $ & $6.55 \times 10^4 $ \\
			\cline{1-6}
			\textbf{Discriminator /} & Parameters & $1.44 \times 10^4$ & $1.56 \times 10^4$ & $1.56 \times 10^4$ & $4.77 \times 10^6$ \\
			\cdashline{2-6}
			\textbf{Environmental model} & FLOPs & $7.35 \times 10^7$ & $1.47 \times 10^8$ & $1.47 \times 10^8$ & $5.95 \times 10^9$\\
			\cline{1-6}
			\multirow{2}*{\textbf{MCTS tree}} & Parameters & - & - & - & $3.07*10^3$ \\
			\cdashline{2-6}
			& FLOPs & - & - & - & -\tnote{1}\\
			\cline{1-6}
			\multirow{2}*{\textbf{Overall}} & Parameters & $1.77 \times 10^5$ & $ 2.60 \times 10^6 $ & $ 2.60 \times 10^6 $ & $4.77 \times 10^6$\\
			\cdashline{2-6}
			& FLOPs & $1.08 \times 10^{10}$ & $4.17 \times 10^8$ & $4.17 \times 10^8$ & $5.95 \times 10^9$\\
			\thickhline
			\cline{1-6}
		\end{tabular}
		\begin{tablenotes}
			\item[1] The Flops of MCTS tree are composed of low-complexity calculations such as add, subtract, multiply and divide. Therefore, its FLOPs is negligible compared to other matrix operations based modules'.
		\end{tablenotes}
	\end{threeparttable}
\end{table*}

In this section, we discuss the complexity of MCTSteg. Specifically, both computational cost and time cost are taken into discussion. And please note that MCTSteg is different from the GAN(Generative Adversarial Network) structure based steganographic schemes. It is composed of an environmental model and a MCTS tree.

For the computational cost, here we take two metrics into consideration, i.e. parameters and~FLOPs(floating point operations). The results are shown in Tab.~\ref{Table9}. We can see that most of the parameters of MCTSteg belong to its environmental model, i.e. SRNet. It is because that the search procedure of MCTSteg is based on a MCTS tree. And its parameters are composed of only hundreds of tree nodes which have been introduced in Sect. ~\ref{SecOverall}. But, on the other hand, SRNet has millions of parameters, i.e. $4.77\times10^6$. As for FLOPs, we can see that the FLOPs of MCTSteg are caused by the environmental model and MCTS tree. In MCTS tree, the FLOPs are composed of low-complexity operations such as add, subtract, multiply, divide, which is negligible compared to the billions of FLOPs caused by environmental model, i.e. $5.95\times10^9$. 

From Tab. \ref{Table9}, it can be seen that the overall FLOPs of MCTSteg are less than those of ASDL-GAN. But for the overall parameters, MCTSteg is higher than others. The reason is that the environmental model in MCTSteg is SRNet while the other schemes adopt the simpler XuNet as environmental model or discriminator. Although SRNet has a higher computational cost, its detection performance is far beyond XuNet's. Therefore, to make the reward function of MCTSteg more precise, we adopt SRNet as our environmental model even though it introduces additional computational cost.

For the time cost in MCTSteg, it is composed of training time and embedding time. Specifically, the training time is spent on training the environmental model i.e. SRNet, which is 67 hour 10 min. For the embedding time, it is decided by two computational budgets, i.e. confidence threshold budget and maximum search count budget. Therefore, we calculate the average time of embedding all images of BOSSBase, which is 3min17sec for MCTSteg, compared with 1.26sec for CMD per image. The longer time of MCTSteg is because that it is a reinforcement learning based framework while CMD is based on a simple non-additive steganographic strategy. However, the most significant issue for steganography is security performance. It has been demonstrated
that no matter which state-of-the-art steganalyzer MCTSteg is against, there is a obvious security performance improvement of MCTSteg compared with other steganographic schemes. Therefore, we think that it is worth for MCTSteg to cause the increment in computational cost and time cost.

\section{Conclusions and Future Work}\label{SecConclu}
In this paper we propose MCTSteg, the first automatic non-additive
steganographic framework based on reinforcement learning paradigm. It
is composed of a MCTS based non-additive steganographer and a
steganalyzer-based environmental model. The major contributions are as
follows:
\begin{itemize}
\item In order to remove the restrictions in existing non-additive
  steganography such as confine of a specific domain and dependence on
  professional knowledge, we have proposed the first reported
  universal automatic non-additive steganographic distortion learning
  framework, which can work in both spatial and JPEG domain. It aims
  at automatically adjusting distortion distribution without human
  intervention. Furthermore, a new distortion adjustment strategy has
  been proposed which helps MCTSteg achieve better security
  performance.
\item To model the game between a non-additive steganographer and a
  target steganalyzer, we have designed fundamental reinforcement
  learning elements including state, action, reward function, and
  environmental model.  Based on the above elements, our proposed
  MCTSteg can effectively combine the search space of MCTS with the
  distortion metric of the underlying distortion minimization
  framework.
\item Extensive experimental results have demonstrated that MCTSteg
  steadily outperforms the state of the art by a clear margin in
  different benchmark datasets, which confirms that a more secure
  distortion adjustment strategy has been learned by MCTSteg.
\end{itemize}

Our future work will focus on the following aspects: (1) we will try
to design a policy network for learning distortion adjustment policies
based on the samples generated by MCTSteg; (2) we will try to increase
the types of actions in MCTSteg and build a more elastic framework.

\ifCLASSOPTIONcaptionsoff
	\newpage
\fi

\bibliographystyle{IEEEtran}
\bibliography{references}

\begin{IEEEbiography}[{\includegraphics[width=1in,height=1.25in,clip,keepaspectratio]{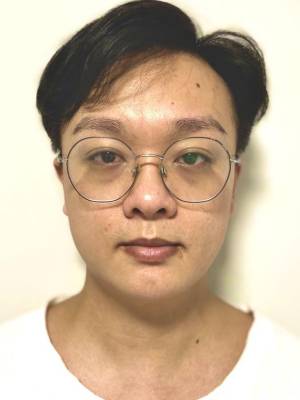}}]{Xianbo
    Mo} received the B.S. degree of computer science and technology from
  Shenzhen University, Shenzhen, China in 2019. He is currently
  pursuing the Ph.D. degree in information and communication
  engineering with Shenzhen University, Shenzhen, China. His current
  research interests include steganography, multimedia forensics and
  reinforcement learning.
\end{IEEEbiography}

\begin{IEEEbiography}[{\includegraphics[width=1in,height=1.25in,clip,keepaspectratio]{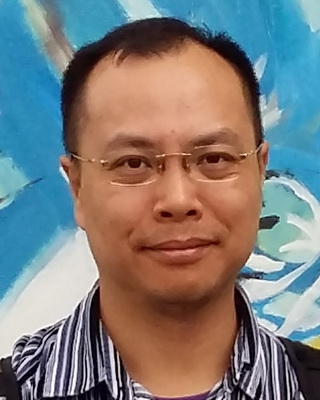}}]{Shunquan Tan (M'10--SM'17)}
  received the B.S. degree in computational mathematics and applied
  software and the Ph.D. degree in computer software and theory from
  Sun Yat-sen University, Guangzhou, China, in 2002 and 2007,
  respectively.

  He was a Visiting Scholar with New Jersey Institute of Technology,
  Newark, NJ, USA, from 2005 to 2006. He is currently an Associate
  Professor with College of Computer Science and Software Engineering,
  Shenzhen University, China, which he joined in 2007. His current
  research interests include multimedia security, multimedia
  forensics, and machine learning.
\end{IEEEbiography} 

\begin{IEEEbiography}[{\includegraphics[width=1in,height=1.25in,clip,keepaspectratio]{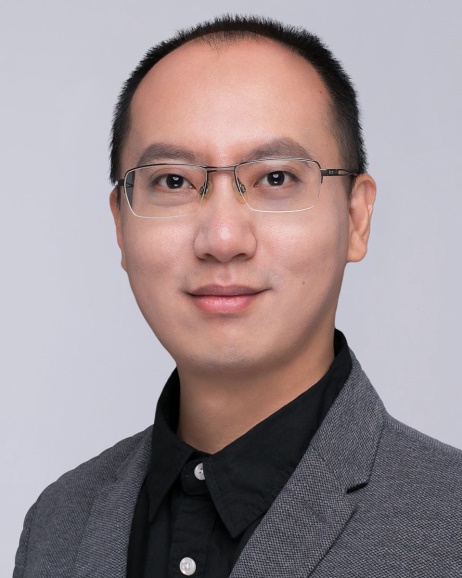}}]{Bin Li (S'07-M'09-SM'17)}
  received the B.E. degree in communication engineering and the Ph.D.
  degree in communication and information system from Sun Yat-sen
  University, Guangzhou, China, in 2004 and 2009, respectively.

  He was a Visiting Scholar with the New Jersey Institute of
  Technology, Newark, NJ, USA, from 2007 to 2008. He is currently a
  Professor with Shenzhen University, Shenzhen, China, where he joined
  in 2009. He is also the Director with the Shenzhen Key Laboratory of
  Media Security and the Vice Director with the Guangdong Key Lab of
  Intelligent Information Processing. He is an Associate Editor of the
  IEEE TRANSACTIONS ON INFORMATION FORENSICS AND SECURITY. His current
  research interests include multimedia forensics, image processing,
  and deep machine learning.
\end{IEEEbiography}

\begin{IEEEbiography}[{\includegraphics[width=1in,height=1.25in,clip,keepaspectratio]{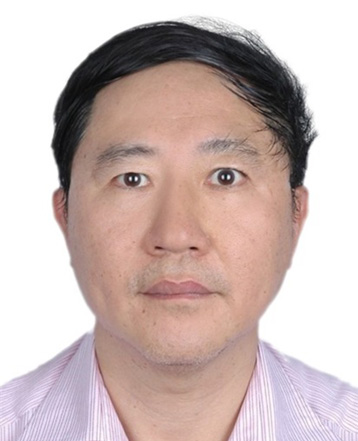}}]{Jiwu Huang (M'98--SM'00--F'16) }
  received the B.S. degree from Xidian University, Xi’an, China, in
  1982, the M.S. degree from Tsinghua University, Beijing, China, in
  1987, and the Ph.D. degree from the Institute of Automation, Chinese
  Academy of Science, Beijing, in 1998. He is currently a Professor
  with the College of Electronics and Information Engineering,
  Shenzhen University, Shenzhen, China. Before joining Shenzhen
  University, he has been with the School of Information Science and
  Technology, Sun Yat-sen University, Guangzhou, China, since
  2000. His current research interests include multimedia forensics
  and security. He is an Associate Editor of the IEEE Transactions on
  Information Forensics and Security.
\end{IEEEbiography}

\end{document}